\documentclass[]{aastex631}

\received{February 20, 2024}
\accepted{June 28, 2024}

\begin{document}

\title{ALMA-IMF XIV: Free-Free Templates Derived from H$41\alpha$ \\ and Ionized Gas Content in Fifteen Massive Protoclusters}

\correspondingauthor{Roberto Galván-Madrid}
\email{r.galvan@irya.unam.mx}

\author[0000-0003-1480-4643]{Roberto Galv\'an-Madrid}
\affiliation{Instituto de Radioastronom\'ia y Astrof\'isica, Universidad Nacional Aut\'onoma de M\'exico, Morelia, Michoac\'an 58089, M\'exico.}

\author[0000-0002-6325-8717]{Daniel J. D\'iaz-Gonz\'alez}
\affiliation{Instituto de Radioastronom\'ia y Astrof\'isica, Universidad Nacional Aut\'onoma de M\'exico, Morelia, Michoac\'an 58089, M\'exico.} 

\author[0000-0003-1649-8002]{Fr\'ed\'erique Motte}
\affiliation{Univ. Grenoble Alpes, CNRS, IPAG, 38000 Grenoble, France.}	

\author[0000-0001-6431-9633]{Adam Ginsburg}
\affiliation{Department of Astronomy, University of Florida, PO Box 112055, USA.}

\author[0000-0003-3152-8564]{Nichol Cunningham}
\affiliation{Univ. Grenoble Alpes, CNRS, IPAG, 38000 Grenoble, France.}
\affiliation{SKA Observatory, Jodrell Bank, Lower Withington, Macclesfield SK11 9FT, United Kingdom.}

\author{Karl M. Menten}
\affiliation{Max-Planck-Institut für Radioastronomie, Auf dem Hügel 69, 53-121 Bonn, Germany.}

\author{Mélanie Armante}
\affiliation{Laboratoire de Physique de l'École Normale Supérieure, ENS, Université PSL, CNRS, Sorbonne Université, Université de Paris, 75005, Paris, France.}
\affiliation{Observatoire de Paris, PSL University, Sorbonne Université, LERMA, 75014, Paris, France.}

\author{Mélisse Bonfand}
\affiliation{Departments of Astronomy and Chemistry, University of Virginia, Charlottesville, VA 22904, USA}

\author{Jonathan Braine}
\affiliation{Laboratoire d'astrophysique de Bordeaux, Univ. Bordeaux, CNRS, B18N, allée Geoffroy Saint-Hilaire, 33615 Pessac, France.}

\author{Timea Csengeri}
\affiliation{Laboratoire d'astrophysique de Bordeaux, Univ. Bordeaux, CNRS, B18N, allée Geoffroy Saint-Hilaire, 33615 Pessac, France.} 

\author{Pierre Dell'Ova}
\affiliation{Laboratoire de Physique de l'École Normale Supérieure, ENS, Université PSL, CNRS, Sorbonne Université, Université de Paris, 75005, Paris, France.}
\affiliation{Observatoire de Paris, PSL University, Sorbonne Université, LERMA, 75014, Paris, France.}

\author[0000-0003-3814-4424]{Fabien Louvet}
\affiliation{Univ. Grenoble Alpes, CNRS, IPAG, 38000 Grenoble, France.}

\author{Thomas Nony}
\affiliation{INAF - Osservatorio Astrofisico di Arcetri, Largo E. Fermi 5, 50125 Firenze, Italy.}
\affiliation{Instituto de Radioastronom\'ia y Astrof\'isica, Universidad Nacional Aut\'onoma de M\'exico, Morelia, Michoac\'an 58089, M\'exico.} 

\author[0000-0002-2162-8441]{Rudy Rivera-Soto}
\affiliation{Instituto de Radioastronom\'ia y Astrof\'isica, Universidad Nacional Aut\'onoma de M\'exico, Morelia, Michoac\'an 58089, M\'exico.} 

\author[0000-0002-7125-7685]{Patricio Sanhueza}
\affiliation{National Astronomical Observatory of Japan, National Institutes of Natural Sciences, 2-21-1 Osawa, Mitaka, Tokyo 181-8588, Japan.}
\affiliation{Department of Astronomical Science, SOKENDAI (The Graduate University for Advanced Studies), 2-21-1 Osawa, Mitaka, Tokyo
181-8588, Japan.}

\author[0000-0003-2300-8200]{Amelia M. Stutz}
\affiliation{Departamento de Astronomía, Universidad de Concepción, Casilla
160-C, 4030000 Concepción, Chile.} 

\author{Friedrich Wyrowski}
\affiliation{Max-Planck-Institut für Radioastronomie, Auf dem Hügel 69, 53-121 Bonn, Germany.}

\author{Rodrigo H. Álvarez-Gutiérrez}
\affiliation{Departamento de Astronomía, Universidad de Concepción, Casilla
160-C, 4030000 Concepción, Chile.} 

\author{Tapas Baug}
\affiliation{S. N. Bose National Centre for Basic Sciences, Sector-III, Salt Lake, Kolkata 700106, India.}

\author{Sylvain Bontemps}
\affiliation{Laboratoire d'astrophysique de Bordeaux, Univ. Bordeaux, CNRS, B18N, allée Geoffroy Saint-Hilaire, 33615 Pessac, France.}

\author{Leonardo Bronfman}
\affiliation{Departamento de Astronomía, Universidad de Chile, Las Condes, 7591245 Santiago, Chile.}

\author{Manuel Fernández-López}
\affiliation{Instituto Argentino de Radioastronom\'\i a (CCT-La Plata, CONICET; CICPBA), C.C. No. 5, 1894, Villa Elisa, Buenos Aires, Argentina.}

\author{Antoine Gusdorf}
\affiliation{Laboratoire de Physique de l'École Normale Supérieure, ENS, Université PSL, CNRS, Sorbonne Université, Université de Paris, 75005, Paris, France.}
\affiliation{Observatoire de Paris, PSL University, Sorbonne Université, LERMA, 75014, Paris, France.}

\author{Atanu Koley}
\affiliation{Departamento de Astronomía, Universidad de Concepción, Casilla
160-C, 4030000 Concepción, Chile.} 

\author{Hong-Li Liu}
\affiliation{ School of physics and astronomy, Yunnan University, Kunming, 650091, P.R. China.}

\author{Javiera Salinas}
\affiliation{Departamento de Astronomía, Universidad de Concepción, Casilla
160-C, 4030000 Concepción, Chile.}

\author{Allison P. M. Towner}
\affiliation{University of Arizona Department of Astronomy and Steward Observatory, 933 North Cherry Avenue, Tucson, AZ 85721, USA.}

\author{Anthony P. Whitworth}
\affiliation{School of Physics and Astronomy, Cardiff University, Cardiff CF24 3AA, UK.}



\begin{abstract}
We use the H$41\alpha$ recombination line to create templates of the millimeter free-free emission in the ALMA-IMF continuum maps, which allows to separate it from dust emission. This method complements spectral-index information and extrapolation from centimeter wavelength maps. We use the derived maps to estimate the properties of up to 34 \textsc{Hii} regions across the ALMA-IMF protoclusters. The hydrogen ionizing-photon rate $Q_0$ and spectral types follow the evolutionary trend proposed by Motte et al. The youngest protoclusters lack detectable ionized gas, followed by protoclusters with increasing numbers of OB stars. The total $Q_0$ increases from $\sim 10^{45}$ s$^{-1}$ to $> 10^{49}$ s$^{-1}$. We used the adjacent He$41\alpha$ line to measure the relative number abundances of helium, finding values consistent with the Galactic interstellar medium, although a few outliers are discussed. A search for sites of maser amplification of the H$41\alpha$ line returned negative results. We looked for possible correlations between the electron densities ($n_e$), emission measures (EM), and $Q_0$ with \textsc{Hii} region size $D$.  
The latter are the better correlated, with $Q_0 \propto D^{2.49\pm0.18}$. 
This favors interpretations where smaller ultracompact \textsc{Hii} regions are not necessarily the less dynamically evolved versions of larger ones, but rather are ionized by less massive stars. Moderate correlations were found between dynamical width $\Delta V_\mathrm{dyn}$ with $D$ and $Q_0$. $\Delta V_\mathrm{dyn}$ increases from about one to two times the ionized-gas sound speed. Finally, an outlier \textsc{Hii} region south of W43-MM2 is discussed. We suggest that this source could harbor an embedded stellar or disk wind.
\end{abstract}



\section{Introduction} \label{sec:intro}

The ALMA-IMF Large Program targeted fifteen of the most massive star-forming clumps within 5 kpc from the Sun \citep{Csengeri17}, based on the larger ATLASGAL sample across the Galaxy \citep{Schuller09,Csengeri14}. The survey description has been presented in \citet{Motte22}, including a proposed evolutionary scheme for the sample based on their activity in the mid- and far-infrared, as well as the presence of hydrogen recombination line emission. The ALMA-IMF consortium has provided science-ready data products to the community, including 12m-array continuum images, together with an imaging and self-calibration pipeline \citep{Ginsburg22}, a combination with single-dish bolometer images  \citep{DiazGonzalez23}, and full-bandwidth line cubes \citep{Cunningham2023}. The two-band (3 mm and 1 mm), two-configuration, mosaic observing mode allows to map the extent of these protocluster clumps with excellent imaging quality \citep{Ginsburg22}. 

The main scientific driver of the project is to measure the Core Mass Function (CMF) in a statistically significant ($> 500$) sample of cores that spans the full range of gas masses, from sub-solar to $\gtrsim 100~M_\odot$ \citep[e.g.,][]{Motte18_NatAs}, something that can only be achieved by observing massive protocluster clumps at different evolutionary stages. 
The more evolved protocluster clumps in the sample have increasingly important feedback by ionisation from ultracompact (UC) and hypercompact (HC) \textsc{Hii} regions \citep[e.g.,][]{Kurtz94,Churchwell02,Hoare07,Purcell13,Klaassen18}.  
The corresponding free-free continuum emission can be an important ``contaminant'' for the otherwise dust-only ALMA millimeter continuum maps. The separation of dust and free-free contributions is therefore a necessary step for the 
correct interpretation of the origin of the emission and derived quantities \citep[e.g.,][]{GM09,Zhang22}.
First results on the evolution of the CMF in young 
protoclusters mostly devoid of \textsc{Hii} regions are presented in \citet{Pouteau22, Pouteau23} and \citet{Nony2023}.  
The CMF of the evolved protocluster G012.80 is presented in \citet{Armante24}, who used the results presented in this paper. 
The CMFs of the global sample, and also as a function of evolution class, are reported in Louvet et al. (subm.). 

ALMA-IMF has been designed to include tools 
to separate the free-free contribution. 
The first one is the use of spectral index information 
\citep{Pouteau22,Nony2023,DiazGonzalez23}. 
However, it is desirable to have an alternative marker that also provides an estimation of the free-free contribution in intensity units, i.e., as a map. Recombination lines are  useful tools to detect and characterize the \textsc{Hii} regions produced by massive star and cluster formation \citep[e.g.,][]{Rugel19,Anderson21}. Lines in the (sub)millimeter are brighter than their centimeter counterparts and mostly free from collisional (``pressure'') broadening, thus they are useful to characterize the denser \textsc{Hii} regions 
\citep[e.g.,][]{KZK08,Quang17,Kim17}. Moreover, their brightness has been shown to be globally consistent with local thermodynamic equilibrium (LTE) in massive star formation regions \citep{Kim18}. Therefore, the velocity-integrated line intensity can be used to estimate the underlying free-free continuum \citep[e.g.,][]{Liu19}. 
This method offers several advantages, namely: covering the whole survey, having approximately the same uv-coverage and beamsize as the corresponding continuum images, and providing the estimate of the free-free contribution directly at a similar frequency  to the continuum. The latter decreases the potential effect of assuming a wrong free-free spectral index, as could be the
case if a extrapolation from centimeter wavelength data is employed. However, the method is not without its caveats, mainly that the line can be out of LTE under certain conditions \citep[e.g.,][]{Walmsley90,Peters12}, and molecular line contamination.   

In this paper, we propose to use the H$41\alpha$ line data within the same ALMA-IMF data set.
In Section \ref{sec:data}  we describe the data.  In Section \ref{sec:ff-estimations} we explain the procedure to obtain the free-free estimation and error maps, the products that we make available, and a validation of the procedure and its limitations. In Section \ref{sec:hiis} we derive physical properties for the \textsc{Hii} regions in the ALMA-IMF protoclusters. In Section \ref{sec:line_params} we use the hydrogen and helium $41\alpha$ line parameters to derive relative He abundances in the ionized gas. In Section \ref{sec:disc} we discuss the benefits and caveats of our proposed technique, look for candidates of recombination line masers, and further discuss the evolution of photoionization within the ALMA-IMF sample. 

\section{Data} \label{sec:data}

For each of the 15 ALMA-IMF protoclusters, we created cubes that cover the $41\alpha$ recombination lines of hydrogen (H$41\alpha$), helium (He$41\alpha$), and carbon (C$41\alpha$). The rest frequencies of the targeted lines are respectively  92.034434, 92.071938, and 92.080355 GHz, which correspond to velocity offsets of the He and C lines with respect to H41$\alpha$ of $-122.165$ and $-149.583$ km s$^{-1}$.
All the cubes were made covering a total width of 270 km s$^{-1}$. The reference velocities for the cube 
construction were set to the LSR systemic velocities of each protocluster reported in Table 1 of \citet{Motte22}.
The cubes were created in \texttt{CASA} version 6.3.0 \citep{CASA2022} using the ALMA-IMF data pipeline \citep{Ginsburg22}. We followed the procedures developed for the imaging of cubes as described in \citet{Cunningham2023}. 

Each cube is cleaned down to a residual level
of $\times 3$ to $\times 5$ the rms noise ($\sigma_\mathrm{rms}$) of channels with H$41\alpha$ emission. This is defined using the \texttt{threshold} parameter of the \texttt{tclean} task.  
Also, a \texttt{startmodel} is used based on the continuum images published in \citet{Ginsburg22}. This permits \texttt{tclean} to run significantly faster because the Fourier Transform of the \texttt{startmodel} is subtracted from the visibilities at the beginning of the cleaning process \citep[see][]{Cunningham2023}. 
Multi-scale cleaning is used by setting the \texttt{deconvolver} and \texttt{scales} parameters in \texttt{tclean}. The scales are selected to clean model components of size zero  (point source), the beamsize, and multiples of $\times 2$ the beamsize until reaching $\sim 8$ arcsec, or about half of the largest recoverable scale in the data \citep{DiazGonzalez23}.  
The \texttt{gain} parameter was set to 0.08 for all the cubes, which is slightly slower than the default (0.1) in \texttt{CASA}. This can be afforded for the case of line-specific cubes.
Finally, the JvM correction \citep{JvM95} is applied to the \texttt{CASA}-generated cubes in order to correct the flux scale of the residuals and of the restored image due to the fact that the volumes of the clean and dirty beams are not equal.\footnote{The scripts to apply the JvM correction that were used for the imaging presented in this paper and in \citet{Cunningham2023} can be found in \url{https://github.com/radio-astro-tools/beam-volume-tools}}     
The cleaning parameters are available in the \texttt{imaging\_parameters.py} script in the public repository of the ALMA-IMF data processing pipeline.\footnote{\url{https://github.com/ALMA-IMF/reduction}}

\section{Free-free estimations} \label{sec:ff-estimations}

\subsection{Procedure} \label{sec:proc}

We use the H$41\alpha$ line to create maps of the free-free emission at the frequencies of the ALMA-IMF continuum images in Bands 3 and 6. 
Using a recombination line within the ALMA-IMF data sets ensures to have a similar $(u,v)$ coverage and angular resolution as the continuum images. 

The free-free estimations are useful to the main scientific case of the ALMA-IMF project of characterizing the dust emission of the dense-core population in the target protoclusters. For this science case the free-free component is mainly considered as a contaminant of the dust continuum emission, but it can also be used to estimate the properties of the 
\textsc{Hii} regions produced by the young ($\sim$ ZAMS) stellar population.  

\begin{figure}
\centering 
\includegraphics[width=0.4\columnwidth]{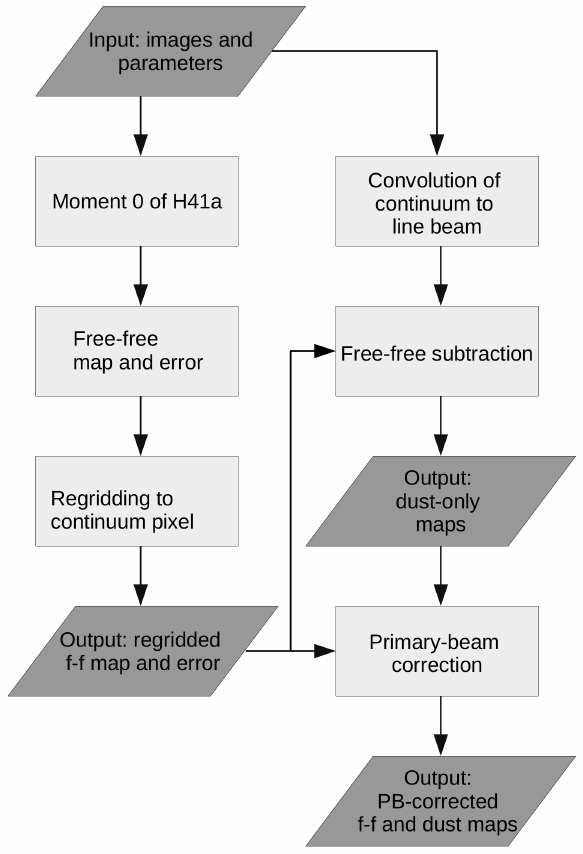}
    \caption{Flowchart of the procedure to create two types of output in each of the observed ALMA continuum bands: free-free maps with their associated errors, and estimation maps of dust-only emission. The primary-beam corrected versions of the previous two are calculated at the end of the procedure.}
    \label{fig:flowchart}
\end{figure}

Figure \ref{fig:flowchart} summarizes the procedure that we execute for each protocluster and band. 
First, three input images for a given protocluster are read in: the H$41\alpha$ line cube and one  continuum image for each band.
Then, a velocity-integrated intensity (moment 0) map $\sum I_{\mathrm{H}41} \, \Delta \mathrm{v}$ in units of Jy beam$^{-1}$ km s$^{-1}$ is created for each field. The velocity range is optimized by visual inspection to cover only channels where the line is detected. A default value of $\pm 10$ km s$^{-1}$ from the cloud systemic velocity is used for the protoclusters where the line is faint or non-detected (see below). 
After this, the free-free estimation map $I_\mathrm{ff}$ in units of Jy beam$^{-1}$ is constructed using the following equation \citep[e.g.,][]{Wilson09,Wenger19}: 

\begin{equation}
I_\mathrm{ff} = 1.432 \times 10^{-4} \bigl( \nu_0^{-1.1} \, T_e^{1.15} \bigr ) \bigl( 1+N_\mathrm{He}/N_\mathrm{H} \bigr ) \sum I_{\mathrm{H}41} \, \Delta \mathrm{v}, 
\label{eq:ff_map}
\end{equation}

\noindent 
where the electron temperature $T_e = 7000$ K, the helium to hydrogen number ratio $N_\mathrm{He}/N_\mathrm{H} = 0.08$, and $\nu_0$ is the central frequency of the H$41\alpha$ line. 
Equation \ref{eq:ff_map} assumes LTE conditions, and we found a posteriori that this assumption is 
generally valid (see Section \ref{sec:LTE-validity}).
$T_e$ in \textsc{Hii} regions is expected to have values from $\gtrsim 5000$ K to $\approx 10^4$ K \citep{Osterbrock89}. Our selection is close to the middle of this range and minimized regions of oversubtraction of the free-free emission in the continuum maps. The selected value of $N_\mathrm{He}/N_\mathrm{H}$ is appropriate for the Milky Way ISM \citep{MendezDelgado20}, and later we show that the \textsc{Hii} regions in our sample are mostly  consistent with it (see Section \ref{sec:line_params}). Further updates in $N_\mathrm{He}/N_\mathrm{H}$ only change $I_\mathrm{ff}$ by a few percent (see Eq. \ref{eq:ff_map}). We consider a $20\%$ error in $T_e$ and propagate it through our measurements to acknowledge the possibility of spatial variations in this quantity. In many of the derived quantities this uncertainty is dominates the errors. 

At millimeter wavelengths it is expected that free-free emission will be optically thin (spectral index $\alpha_\mathrm{ff} = -0.1$) in all \textsc{Hii} regions except possibly in the densest ones \citep[e.g.,][]{Keto02,GM09,Rosero19}. The actual detection of H$41\alpha$ guarantees that the underlying free-free continuum is not opaque, otherwise the line would be absorbed. 
For a homogeneous \textsc{Hii} region, the transition or turn-over frequency between the optically thick and thin regimes is in the range from 50 GHz to 60 GHz for an emission measure EM (see Section \ref{sec:hiis}) as high as $10^{10}$ cm$^{-6}$ pc \citep{Kurtz05}, although \textsc{Hii} regions with density gradients could be partially optically thick ($\alpha_\mathrm{ff} \sim 1$) at frequencies $\gtrsim 100$ GHz \citep{KZK08,GM09}.    
In our procedure, the output free-free map at the frequency of H$41\alpha$ (92.03 GHz) is rescaled to the desired frequency of the continuum image assuming $\alpha_\mathrm{ff} = -0.1$. 
The flux scalings to the Band 3 (98.6 GHz) and Band 6 (224.6 GHz) continuum frequencies are then 0.99 and 0.92, respectively. 
If the correct free-free spectral index in those spots were $\alpha=1$ all the way to Band 6, the continuum scaling factors should instead be 1.07 and 2.44. 
It is possible that in Band 3 the optically-thin  assumption breaks 
in a few compact spots with very dense \textsc{Hii} regions, but this is highly unlikely in Band 6. 
 
Maps of the error of the free-free intensity $\sigma_\mathrm{ff}$ are also generated. These are approximated by the noise in the moment 0 map $\sigma_\mathrm{m0}$ and the error in the assumed electron temperature $\sigma_{T_e}$ as in: 
\begin{equation} \label{eq:err_ff}
\sigma_\mathrm{ff} \approx I_\mathrm{ff} \Bigl[ \Bigl( \frac{\sigma_\mathrm{m0}}{\sum I_{{\rm H}41} \Delta \mathrm{v}} \Bigr)^2 + \Bigl( \frac{\sigma_{T_e}}{T_e} \Bigr)^2 \Bigr]^{0.5}.  
\end{equation}

The $I_\mathrm{ff}$ and $\sigma_\mathrm{ff}$ maps are regridded to the pixel geometry of the input continuum images, which in turn are convolved to the slightly larger 
beamsize \citep[see][]{Cunningham2023} of the line cube and moment 0 maps. A subtraction is then performed of the convolved continuum image minus the regridded free-free map, resulting in an estimated map of pure-dust emission $I_\mathrm{dust}$.
Finally, the primary-beam (PB) correction is applied to the regridded free-free estimation map and the associated error map, as well as to the pure dust map. Performing the PB correction as a last step has the advantage of working with images with spatially uniform  noise levels 
in the intermediate steps. 

\begin{table}[ht!]
	\centering
    \small 
	\begin{tabular}{cccc} 
		\hline
		Protocluster & HPBW, PA & [$v_\mathrm{min}$,$v_\mathrm{max}$] 
		& $\sigma_\mathrm{m0}$ \\
		& [$\arcsec \times \arcsec,~\deg$] & 
		[km s$^{-1}$] & 
		[mJy beam$^{-1}$ km s$^{-1}$] \\
		\hline
		G008.67 & $0.95\times0.66$, 66.7 & [6.8,84.0] &  20 \\
		G010.62 & $0.57\times0.45$, $-67.9$ & [-39.1,41.9] & 14 \\
		G012.80 & $2.28\times1.93$, 84.9 & [-5.8,84.4] &  83 \\
		G327.29 & $0.71\times0.61$, 57.2 & [-63.4,-19.2] & 9 \\
		G328.25$\dagger$ & $0.99\times0.93$, 75.0 & [-53.0,-33.0] & 9 \\
		G333.60 & $0.78\times0.74$, 28.2 & [-103.3,5.2] & 19 \\
		G337.92 & $0.80\times0.75$, $-77.5$ & [-52.8,-8.7] & 9 \\
		G338.93$\dagger$ & $0.66\times0.63$, 4.7 & [-72.0,-52.0] & 5 \\
		G351.77 & $2.20\times1.90$, 89.8 & [-29.0,20.6] & 42 \\
		G353.41 & $2.18\times1.85$, 73.6 & [-46.7,19.5] & 55 \\
		W43-MM1$\dagger$ & $0.92\times0.52$, $-81.9$ & [80.3,115.3] & 6 \\
		W43-MM2 & $0.43\times0.33$, $-77.8$ & [61.9,122.6] & 5 \\
		W43-MM3 & $0.61\times0.43$, $-84.0$ & [61.1,120.0] & 7 \\
		W51-E & $0.39\times0.35$, $-73.0$ & [22.8,100.0] & 11 \\
		W51-IRS2 & $0.41\times0.37$, $-68.8$ & [8.7,93.4] & 11 \\ 
		\hline
	\end{tabular}
    \centering 
    \caption{{\small H$41\alpha$ cube and moment 0 parameters. Columns: (1) protocluster name; (2) cube half-power beamwidth and position angle; (3) visually determined velocity range of the moment 0 integration (a default $\pm 10$ km s$^{-1}$ is used for the non-detections marked with a $\dagger$); (4) MAD standard deviation of the moment 0 image.}}
	\label{tab:H41_params}
\end{table}

\subsection{Products} \label{sec:products}

Figure \ref{fig:ff-maps} shows the PB-corrected free-free estimation maps for the 12 ALMA-IMF protoclusters with a detection of H$41\alpha$ emission. 
Figure \ref{fig:comparison-maps-app} in Appendix \ref{app:pure-dust} gives a comparison between the original continuum maps, the free-free estimations, and the pure-dust maps at 1.3 mm. 

The free-free estimation is only performed in pixels of the moment 0 maps above a threshold of $5\sigma$, where $\sigma$ is defined as the robust standard deviation using the median absolute deviation (MAD) implemented in \texttt{astropy.stats.mad\_std}. 
Table \ref{tab:H41_params} lists the basic parameters for the preparation of the free-free estimation images. 
Two protoclusters were clear non-detections in H$41\alpha$ (G338.93 and W43-MM1). One protocluster  (G328.25) was a tentative detection, but  further  inspection showed the spectrum was dominated by molecular-line contamination. 
The free-free estimation maps of these three  protoclusters originally contained a few pixels with values larger than zero due to molecular lines in the moment 0 integration, therefore, we manually set their free-free maps to zero. 

The data products are further described in Appendix \ref{app:products}. 
The scripts associated with this paper\footnote{\url{https://github.com/ALMA-IMF/h41a-freefree}} permit the user to easily recalculate any free-free estimation image by editing \texttt{ffsub.py}, which makes use of the functions defined in \texttt{ff\_tools.py}. The code can also be easily applicable to other data sets \citep[e.g.,][for ALMA observations of the W49N protocluster with observations of H$30\alpha$]{Nony24}. 
We caution the user against blindly using the calculated free-free templates in regions where it dominates the total continuum emission, or at the position of line rich hot molecular cores. In the former case, the generated error maps ($\sigma_\mathrm{ff}$) become particularly useful to weigh in the relative contributions of dust and free-free. In the latter, molecular lines can dominate over the recombination line emission. 

\begin{figure*}
\centering
\includegraphics[width=0.48\linewidth]{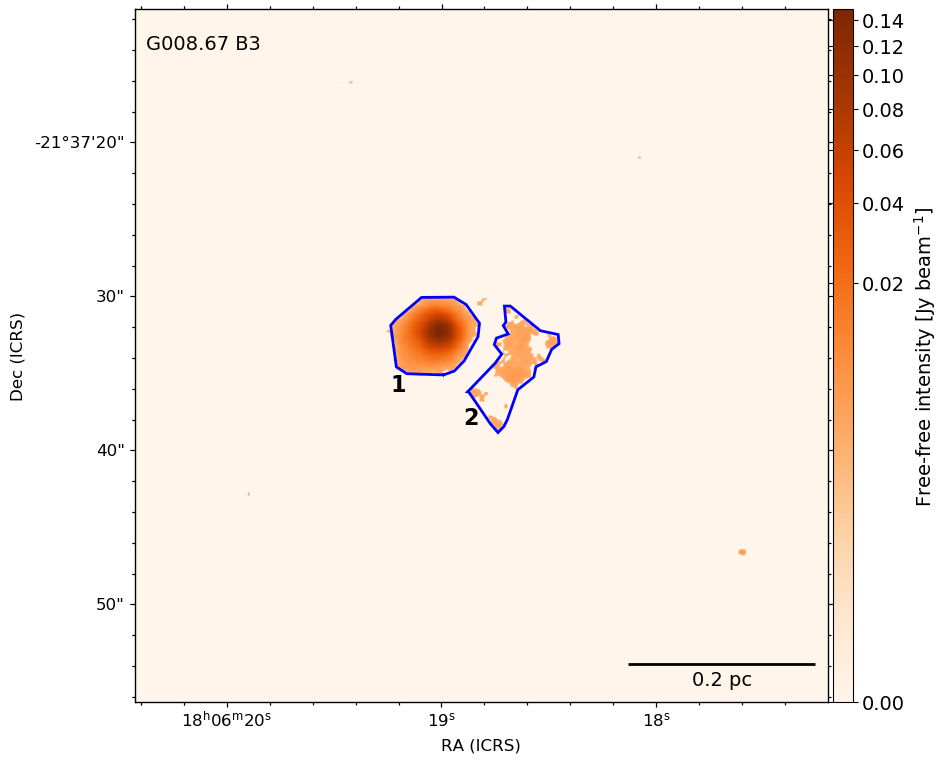}~ 
\includegraphics[width=0.48\linewidth]{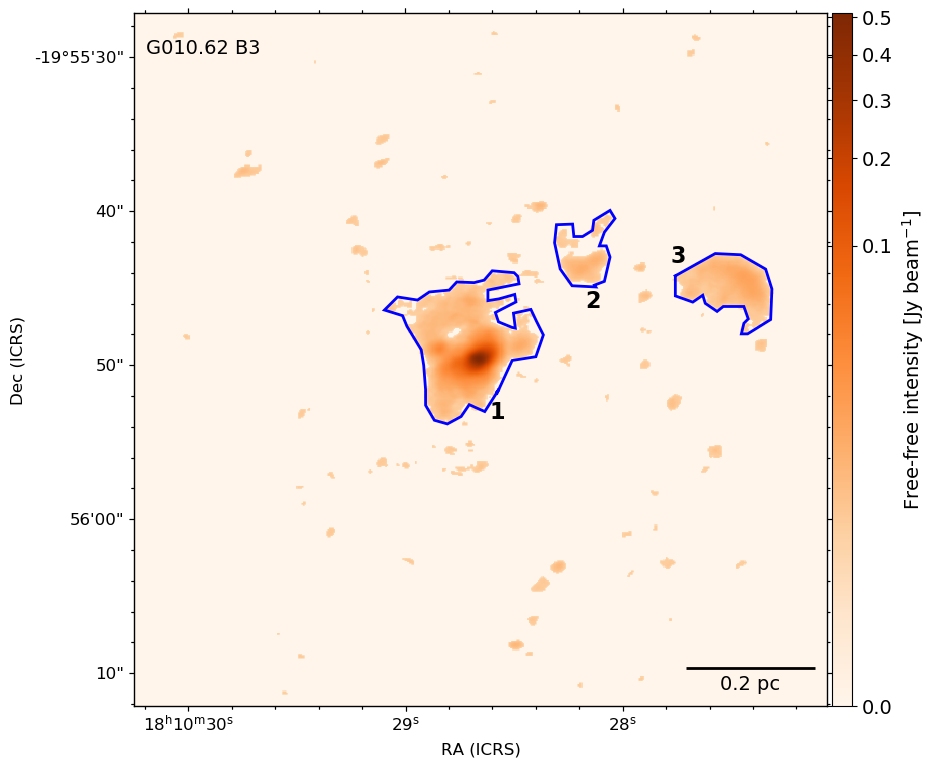}~\\
\includegraphics[width=0.48\linewidth]{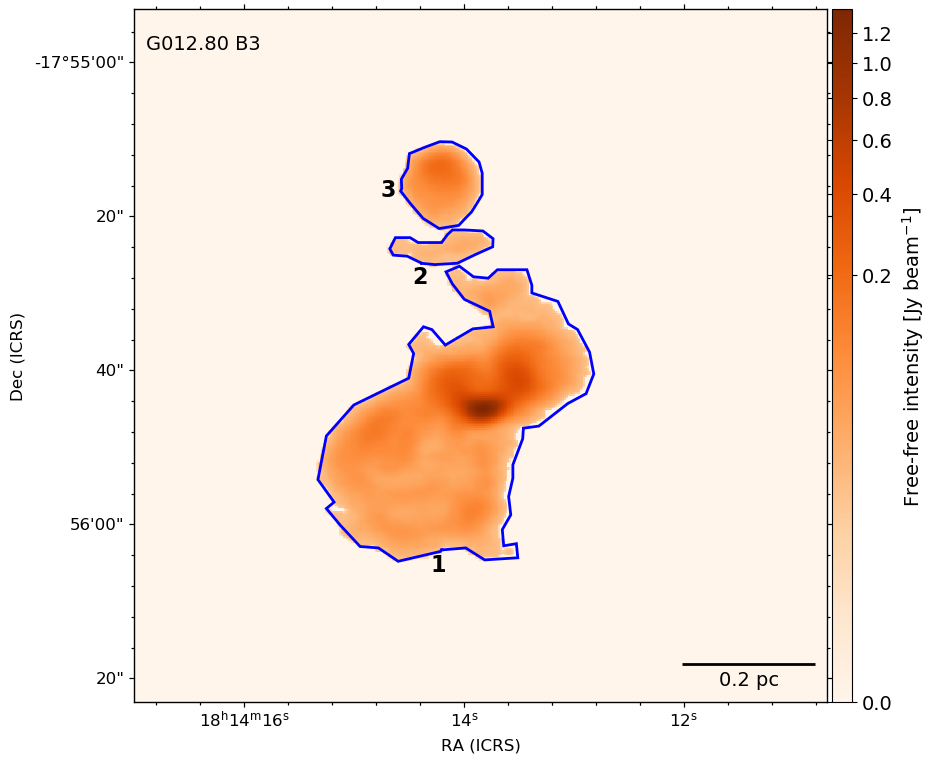}~ 
\includegraphics[width=0.48\linewidth]{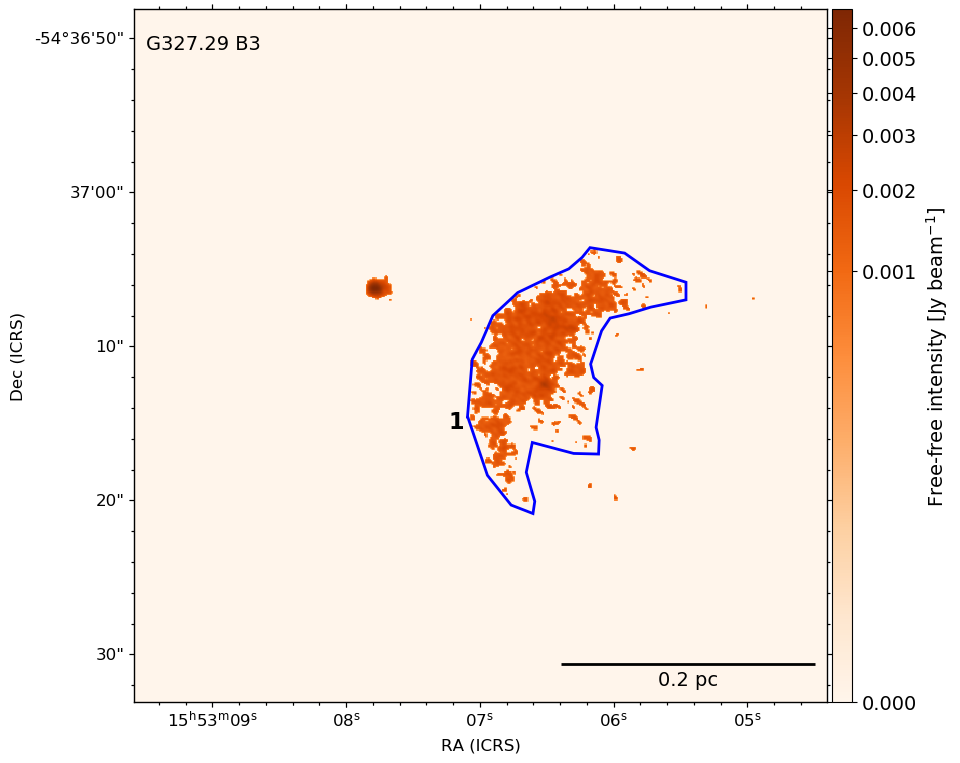}~\\
\includegraphics[width=0.48\linewidth]{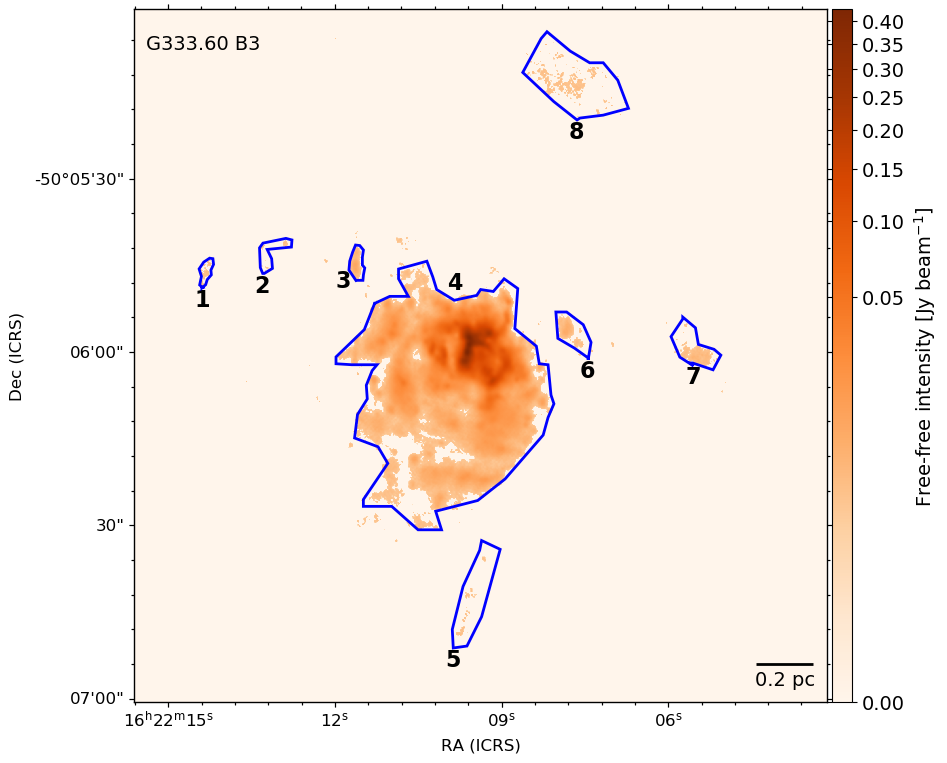}~
\includegraphics[width=0.48\linewidth]{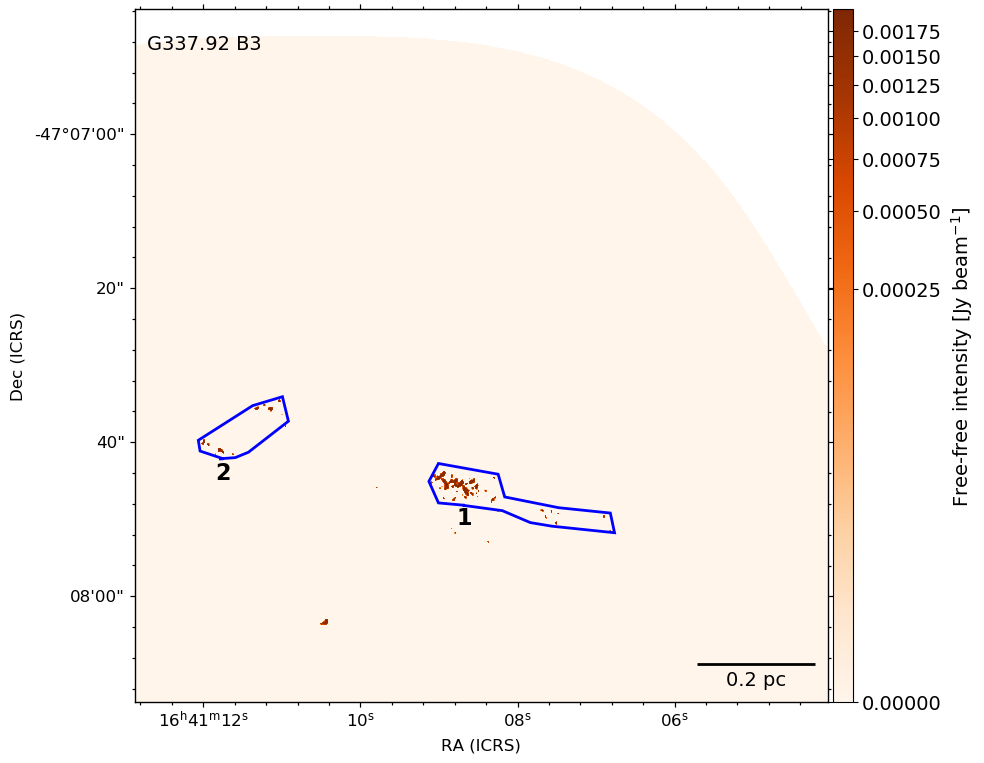}~ 
\caption{Free-free estimation maps in Band 3 for protoclusters G008.67 (top-left), G010.62 (top-right), G012.80 (middle-left), G327.29 (middle-right), G333.60 (bottom-left), and G337.92 (bottom-right). The compact emission at the center of G327.29 is dominated by contamination from molecular lines (see Section \ref{sec:validation}).}
\label{fig:ff-maps}
\end{figure*}

\addtocounter{figure}{-1}
\begin{figure*}
\centering
\includegraphics[width=0.48\linewidth]{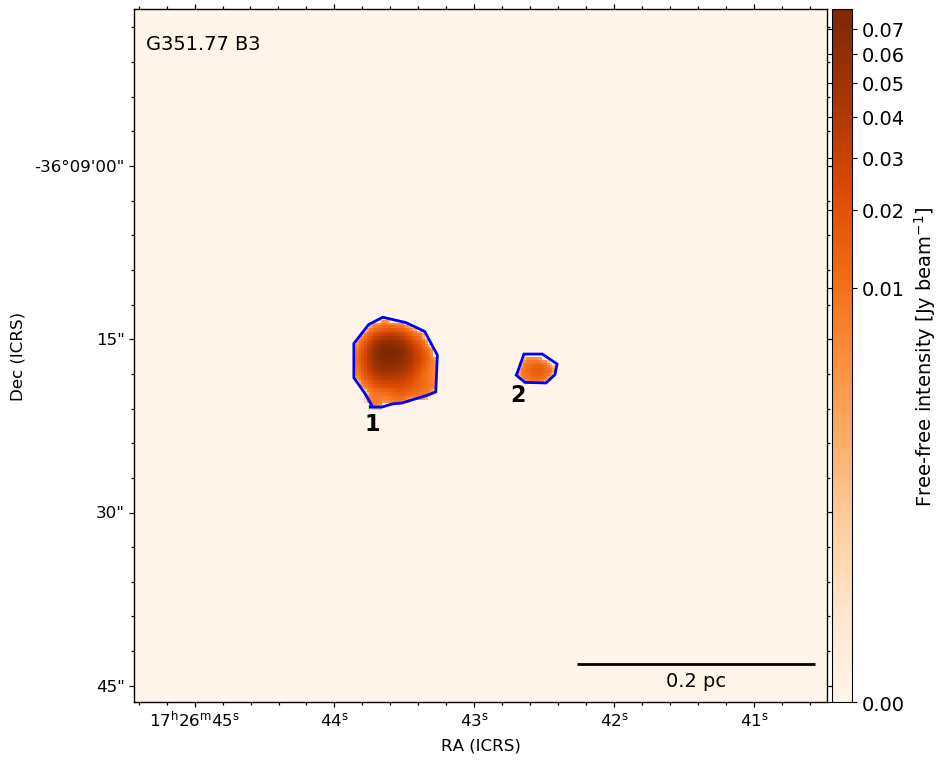}~
\includegraphics[width=0.48\linewidth]{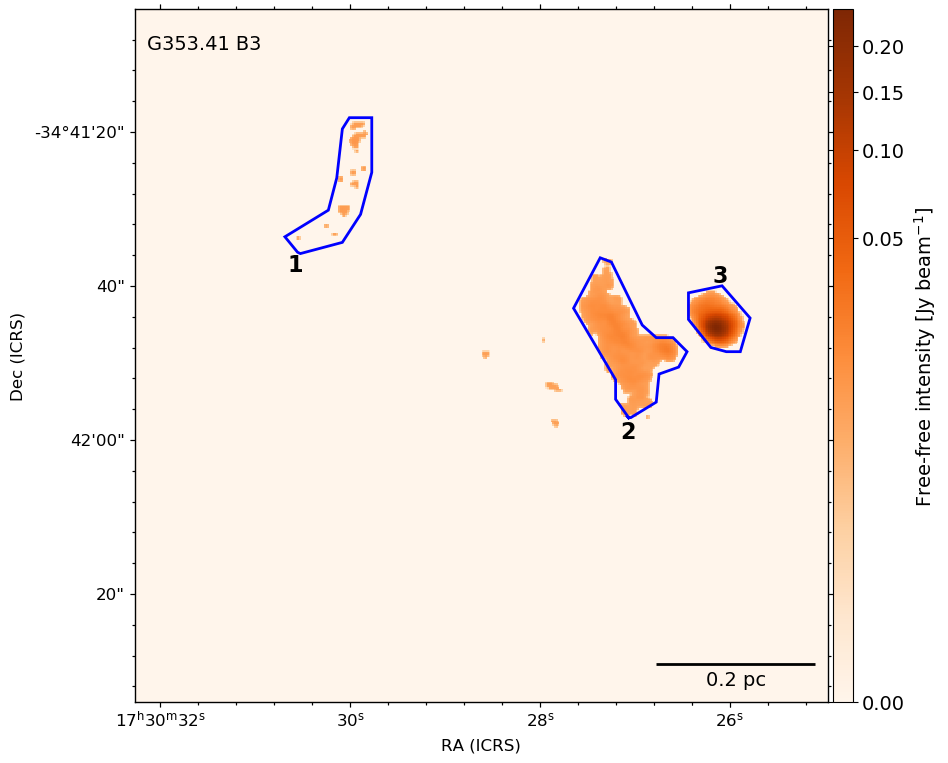}~\\ 
\includegraphics[width=0.48\linewidth]{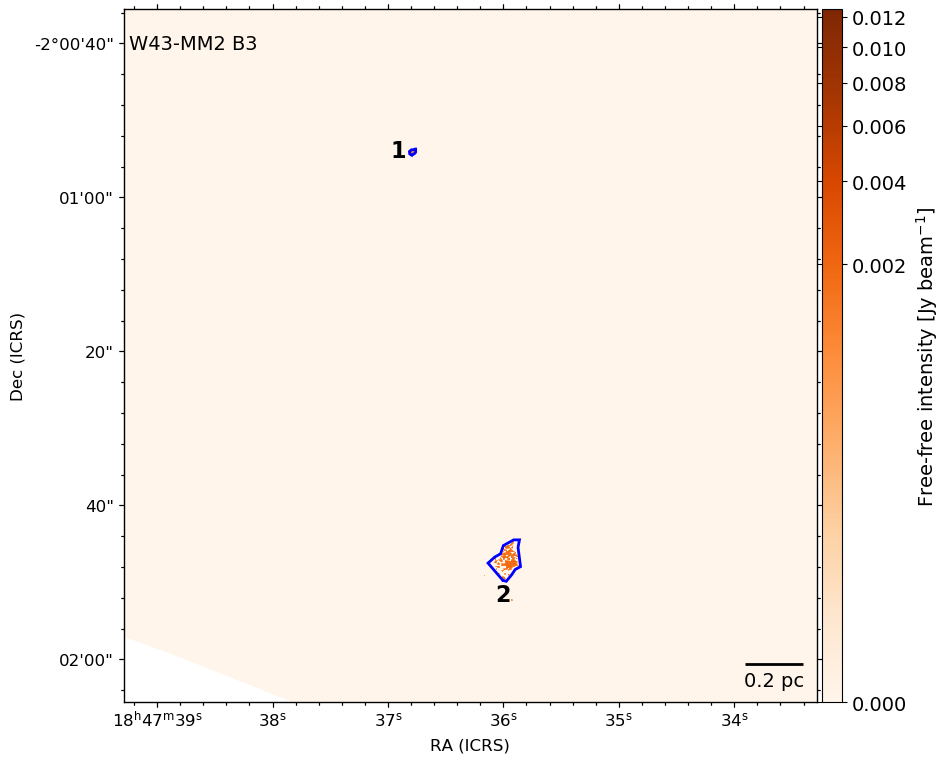}~
\includegraphics[width=0.48\linewidth]{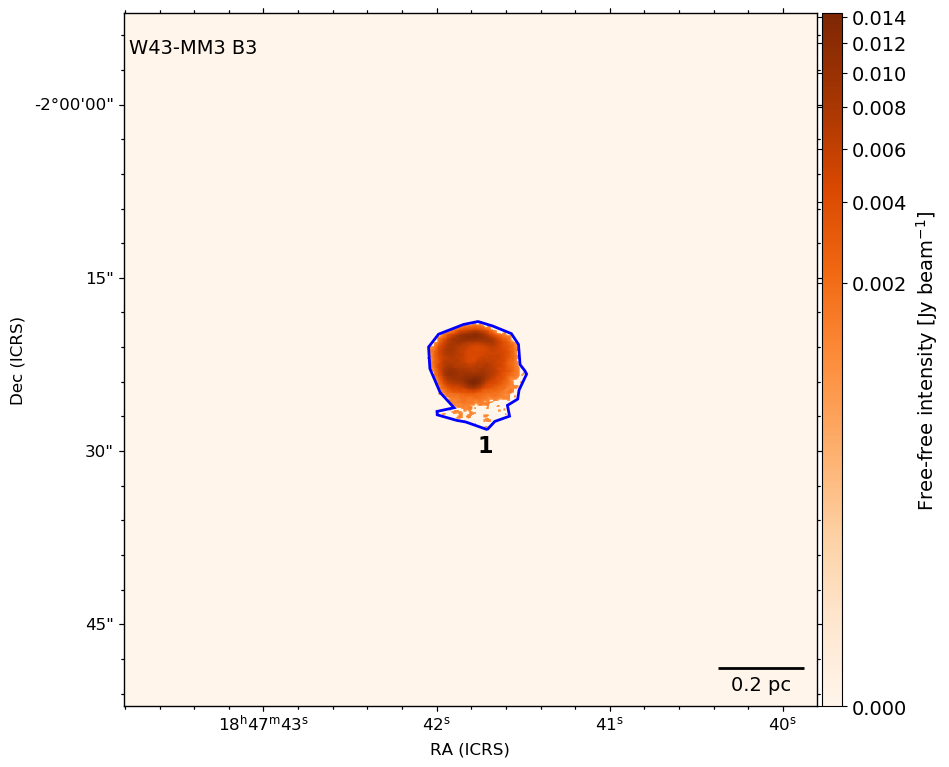}~\\
\includegraphics[width=0.48\linewidth]{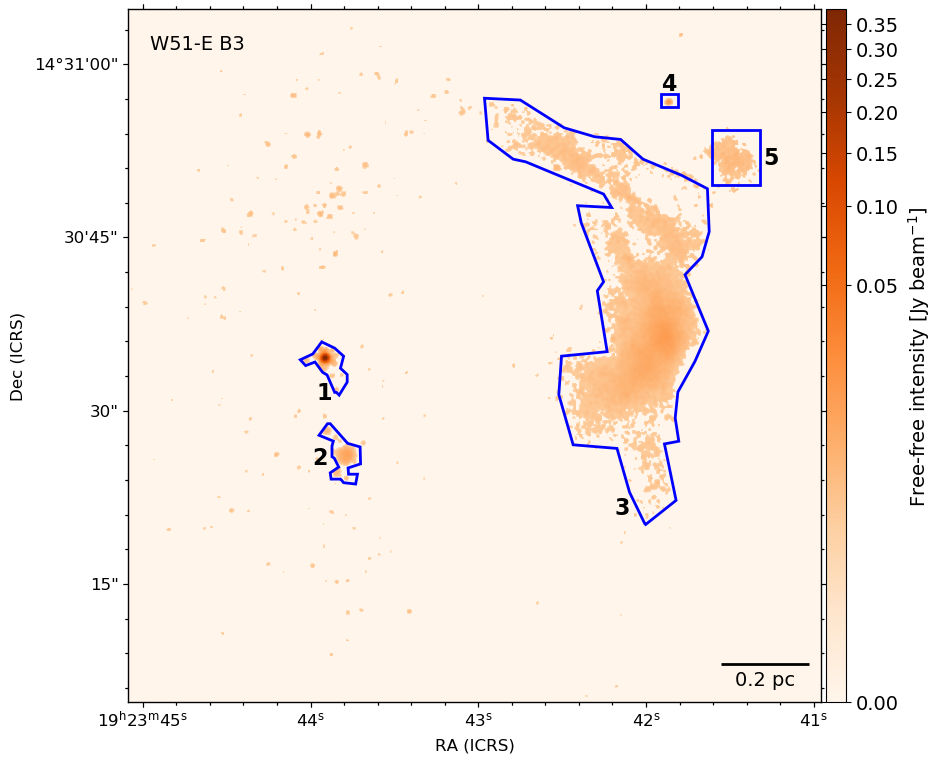}~
\includegraphics[width=0.48\linewidth]{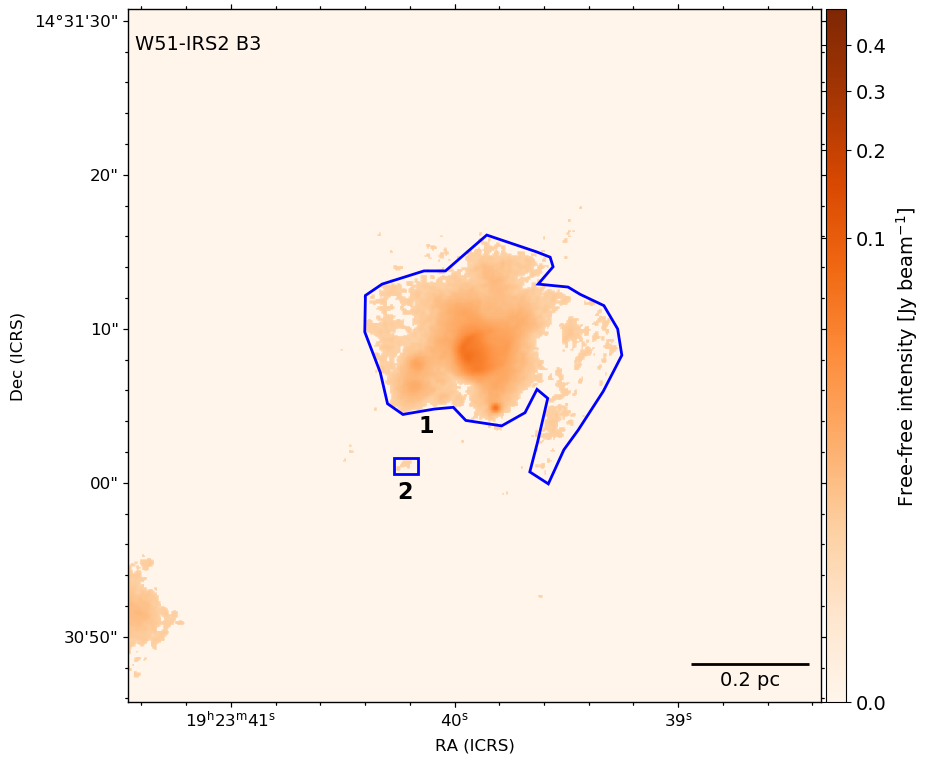}~ 
\caption{Continued. Free-free estimation maps in Band 3 for protoclusters G351.77 (top-left), G353.41 (top-right), W43-MM2 (middle-left), W43-MM3 (middle-right), W51-E (bottom-left), and W51-IRS2 (bottom-right).}
\end{figure*}

\subsection{Validation} \label{sec:validation}

\begin{figure*}
\centering
\includegraphics[width=\textwidth]{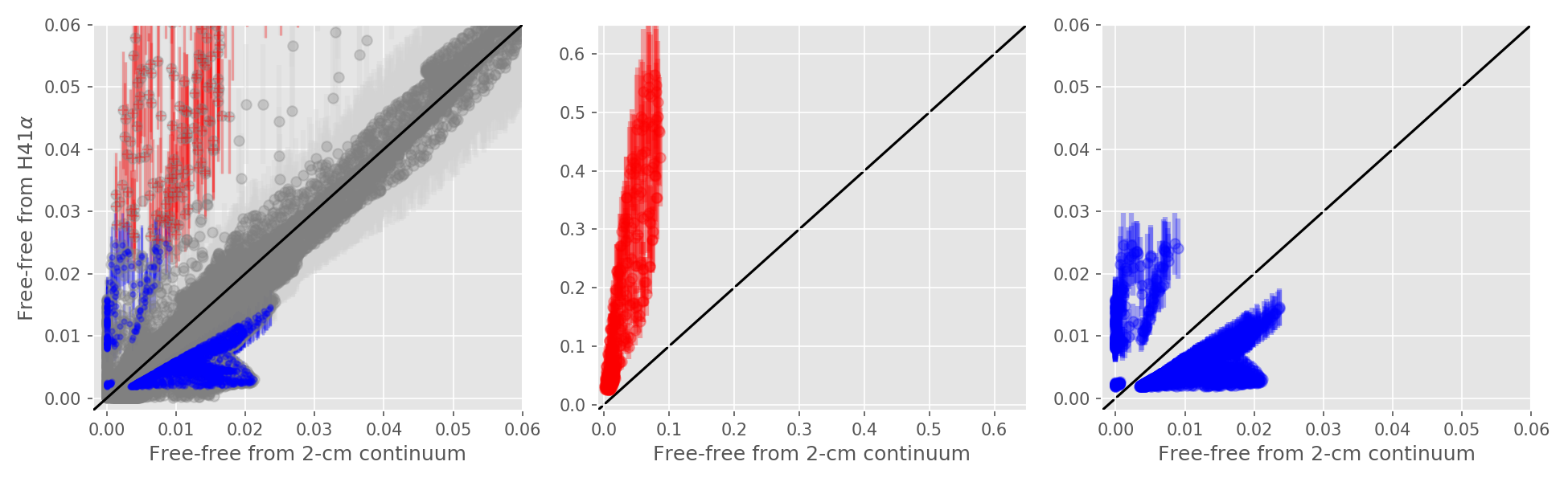}
\caption{Validation of the procedure to estimate the free-free emission in the ALMA-IMF maps via the H41$\alpha$ line. The plots show a pixel-by-pixel comparison between the free-free intensity at 3 mm derived from H$41\alpha$ vs that derived from the 2-cm continuum, assuming a spectral index of $-0.1$. We used the W51-IRS2 and W51-E protoclusters, for which there are also high-quality VLA data at a similar angular resolution \citep{Ginsburg16}. 
The left panel shows all the 145,862 pixels with an intensity above $5\sigma$ in both maps. $96.8\%$ of those pixels fall within $\pm3\sigma$ of the 1-to-1 relation. The middle panel shows the $0.2\%$ of the pixels that are defined as bright outliers; their location is analyzed in Fig. \ref{fig:bright-H41}. The right panel shows the faint outliers ($3\%$ of the pixels), which are mostly in the periphery of \textsc{Hii} regions and have lower signal to noise ratio. The colored points and bars in the middle and right panels have the same color as their error bars in the left panel.}
\label{fig:validation}
\end{figure*}

A typical method to separate the contributions of dust and free-free in  (sub)millimeter continuum maps is to use a centimeter continuum map dominated by free-free and extrapolate it to shorter wavelengths, assuming an appropriate spectral index \citep[e.g.,][]{CG2012,Zhang22}. However, for a survey such as 
ALMA-IMF, using the H41$\alpha$ line within the same data set 
has several advantages, namely: 
i) that the assumed $\alpha_\mathrm{ff} = -0.1$ is more likely to be correct when the scaling is done within millimeter wavelengths, and not from the centimeter to the millimeter, also, the scaling is smaller; ii) that 
the $(u,v)$ coverages of the continuum and line maps are very similar, thus their beamsizes and largest recoverable scales are also quite similar; 
iii) that there are no extra induced flux-calibration or astrometric errors from using different instruments; and iv) that ten out of fifteen ALMA-IMF protoclusters have declinations $< -15^\circ$, which makes them difficult to observe with the VLA, at least  with a good synthesized beam.\footnote{In the ancillary VLA continuum data for the observable ALMA-IMF protoclusters, the synthesized beam is usually larger and always more elliptical than in the ALMA images (Rivera-Soto et al., in prep.).}  Actually, five out of ten protoclusters are unobservable with the VLA.

To cross-validate our method, we compare the H$41\alpha$-derived free-free estimation $I_\mathrm{ff,H41}$ at 98 GHz for the W51 protoclusters to an independent estimation $I_\mathrm{ff,2cm}$ derived from the Ku-band (2.2 cm, 13.4 GHz) VLA image presented in \citet{Ginsburg16}. The beamsize and position angle of the original 2-cm image are $\mathrm{BMAJ}=0.342$ arcsec, $\mathrm{BMIN}=0.332$ arcsec, and $\mathrm{BPA}=14.83$ deg. We convolve this image to the beam of the H$41\alpha$-derived free-free map: 
$\mathrm{BMAJ}=0.408$ arcsec, $\mathrm{BMIN}=0.374$ arcsec, and $\mathrm{BPA}=-68.77$ deg, and regrid it to the same pixel geometry. Next, we assume that the 2-cm free-free emission is optically thin and rescale the cm map by a factor $(98.6/13.4)^{-0.1}$ to estimate the free-free intensity in the ALMA-B3 map. 
Finally, we mask out all pixels below $5\sigma = 5 \times 92~\mu$Jy beam$^{-1}$, where $\sigma$ is defined as in Section \ref{sec:products}.

Figure \ref{fig:validation} shows a comparison of the free-free maps at 98.6 GHz derived using the H$41\alpha$ line and the 2-cm continuum. The comparison is done for those pixels with a valid measurement (i.e., $> 5\sigma$) in both images. 
It is seen that the vast majority of pixels follow the line where the ratio of the estimations $I_\mathrm{ff,H41}/I_\mathrm{ff,2cm} = 1$. 
From 145,862 pixels in the left panel of Fig. \ref{fig:validation}, $96.8~\%$ are within $3\sigma$ of the one-to-one relation, where $\sigma$ is determined for each pixel using the H$41\alpha$ free-free error map. The outlier pixels can be separated into two categories, bright and faint. 
The bright outliers (red points), those pixels with $I_\mathrm{ff,H41} > 0.025$ Jy beam$^{-1}$ and beyond $3\sigma$ from $I_\mathrm{ff,H41}/I_\mathrm{ff,2cm} = 1$, are only 328 pixels, or 4.7 times the beam area. Further inspection reveals that these pixels are  
located exactly at the position of
the two well known UC HII regions: d2 in W51-IRS2 and e2 W51-E \citep[e.g.,][]{Zhang98,Goddi16}. 

\begin{figure*}
\centering
\includegraphics[width=0.48\linewidth]{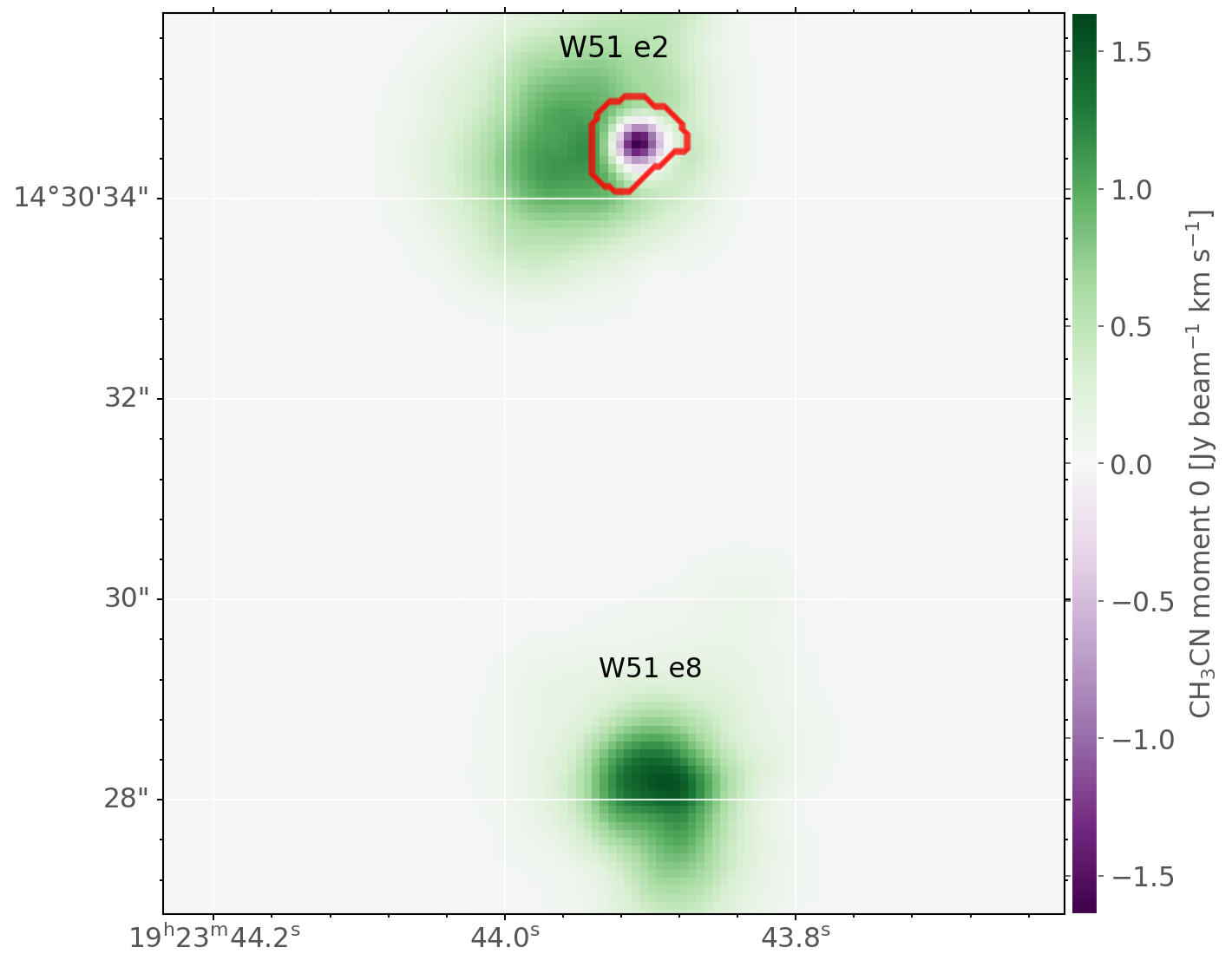}~ 
\includegraphics[width=0.51\linewidth]{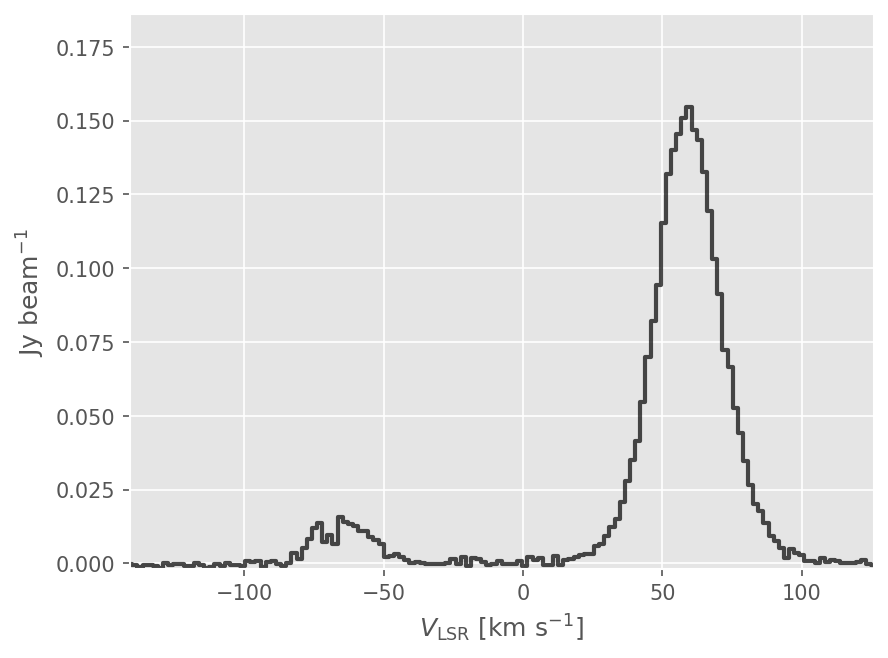}~\\
\includegraphics[width=0.48\linewidth]{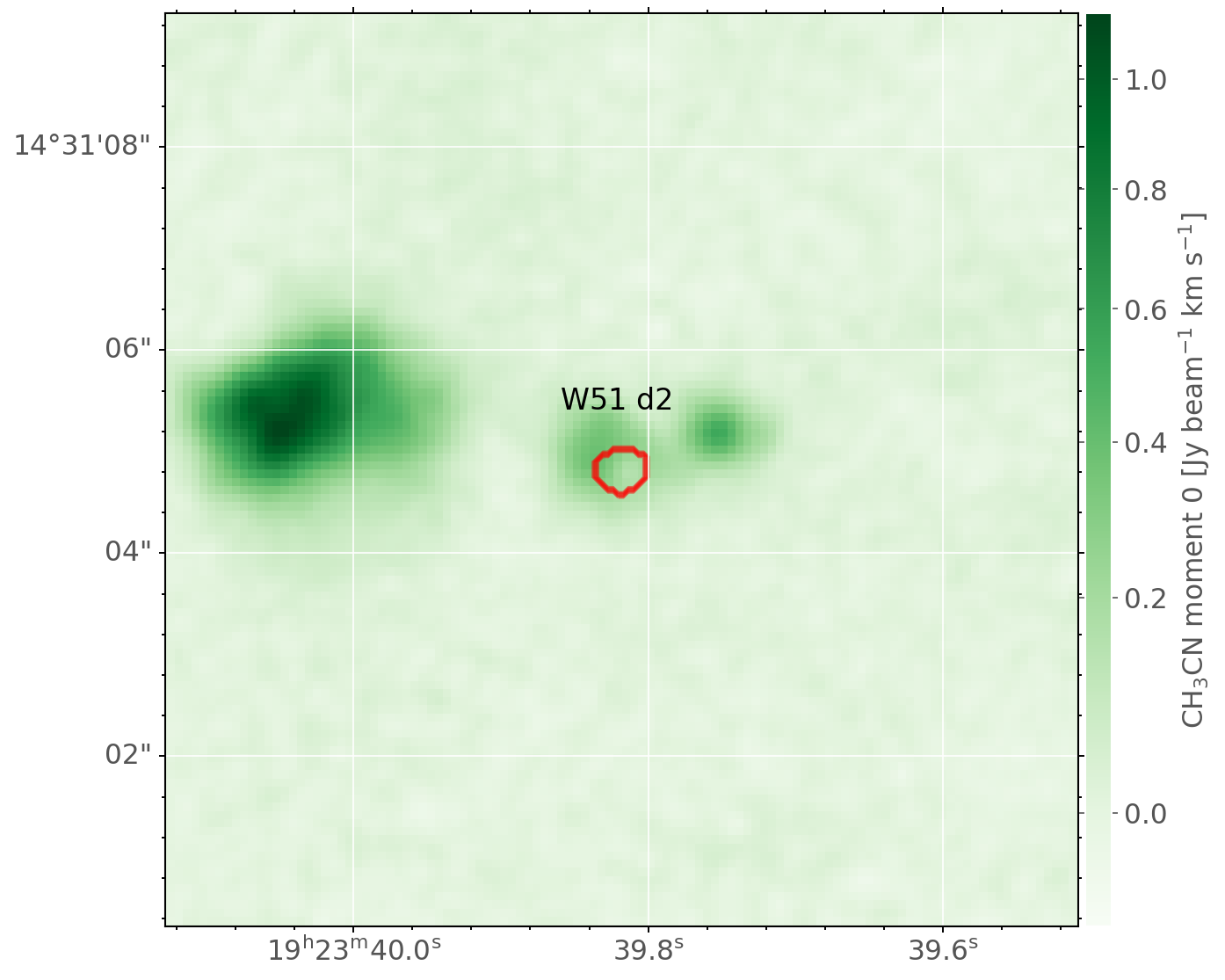}~ 
\includegraphics[width=0.51\linewidth]{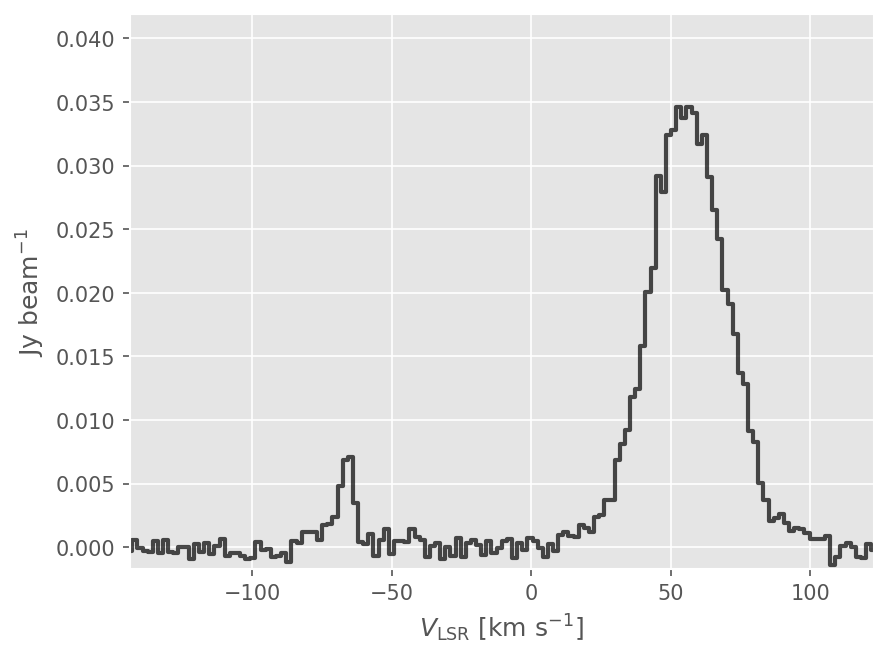}
\caption{Detailed look at the areas labeled as ``bright outliers'' (red points) in Fig. \ref{fig:validation}. The \textit{top-left} panel shows the outliers at the position of the well-known HC \textsc{Hii} region W51 e2. The CH$_3$CN line is seen in absorption at this location due to the high brightness temperature of the partially optically thick \textsc{Hii} emission.  
The \textit{bottom-left} panel shows the location of the outliers at the position of the UC \textsc{Hii} region W51 d2, which is not as bright in the millimeter continuum, therefore CH$_3$CN is not seen in absorption. The right panels plot the corresponding H$41\alpha$ and He$41\alpha$ spectra inside the red contours, showing that molecular-line contamination is not significant. Therefore, the most likely origin of the bright outliers is an underestimation of the free-free extrapolation from cm wavelengths. }
\label{fig:bright-H41}
\end{figure*}

Figure \ref{fig:bright-H41} shows the two previously mentioned regions overlaid on the integrated emission of CH$_3$CN in Band 3. The red contours mark the location of the bright-outliers (central panel of Fig.  \ref{fig:validation}). In the literature, the e2 region (top row of Fig. \ref{fig:bright-H41}) is subdivided into the bright UC HII region e2-west (e2w) and the hot-core emission peaks e2e and e2nw \citep[e.g.,][]{Shi10,Ginsburg17}. The  bright-outlier region matches mostly with e2w. The underlying continuum of this \textsc{Hii} region is bright enough to absorb the CH$_3$CN line. The recombination line  spectrum within this mask (top-right panel of Fig. \ref{fig:bright-H41}) shows a clear detection of the H$41\alpha$ and He$41\alpha$ lines, without evidence of significant contamination from molecular lines. The bottom row of Fig. \ref{fig:bright-H41} shows the second region with bright outliers, which matches the position of the UC HII region W51 d2. The situation is very similar to that of e2, except that the background continuum is not bright-enough to produce absorption of the CH$_3$CN line. 

From the previous analysis, we conclude that the origin of the H41$\alpha$ bright-outlier areas could either be: i) a true yet moderate enhancement of the recombination line compared to LTE conditions; or ii) the optically-thin free-free extrapolation from 13 GHz to 98 GHz being invalid, i.e., having partially optically-thick free-free emission with a spectral index $-0.1 < \alpha_\mathrm{ff} < 2$; or a combination of the two. For the case of W51, the spots of bright outlier H$41\alpha$ emission do not appear to have significant contamination from molecular lines (in W51-E and W51-IRS2 we integrated the velocity ranges [22.8,100] and [8.7,93.4] km s$^{-1}$, respectively,  see Table \ref{tab:H41_params}). We have inspected the correspondence of the 76 sources in the hot molecular core (HMC) catalogue of \citet{Bonfand24_arxiv} with our recombination line maps. Eighteen HMCs overlap with valid pixels in our free-free estimation maps. Analysis of the spectra around their positions shows that recombination line emission dominates in all cases except for G327 HMC 1  and G337 HMCs 1 and 4. These areas are not labeled as \textsc{Hii} regions in Fig. \ref{fig:ff-maps}, and are therefore excluded from our analysis. 

\section{\textsc{Hii} region properties} \label{sec:hiis}

We use the derived 3 mm free-free maps to estimate the average properties of \textsc{Hii} regions in the ALMA-IMF protoclusters. 
For this, we first define apertures of distinct \textsc{Hii} emission in each of the 12 protoclusters with recombination line detection 
(see Section \ref{sec:products}). The apertures are marked in Fig. \ref{fig:ff-maps} and provided in the data release  (\texttt{.reg} files) associated with this paper. We then used the following equations \citep[see, e.g.,][]{Wilson09,RS20} to derive their average emission measures (EM, pc cm$^{-6}$), electron densities ($n_e$, cm$^{-3}$), and hydrogen ionizing-photon rates ($Q_0$, s$^{-1}$): 

\begin{equation} \label{eq:tauff}
\tau_\mathrm{ff} = \ln \biggl [ \biggl ( 1 - \frac{I_\nu}{B_\nu(T_e)} \biggr )^{-1} \biggr ],
\end{equation}

\begin{equation} \label{eq:EM}
\mathrm{EM} = 12.143\tau_\mathrm{ff} \Bigl[ \frac{\nu}{\mathrm{GHz}} \Bigr ]^{2.1}\Bigl [ \frac{T_e}{\mathrm{K}} \Bigl ]^{1.35} =  \int n_e^2 dl,
\end{equation}

\begin{equation} \label{eq:Q0}
Q_0 = \frac{\pi}{6} \alpha_\mathrm{B} n_e^2 D^3 = \frac{\pi}{6} \alpha_\mathrm{B} D^2 \mathrm{EM},   
\end{equation}

\noindent
where $\tau_\mathrm{ff}$ is the free-free continuum optical depth, $I_\nu$ is the average intensity of the pixels within the aperture, $B_\nu(T_e)$ is the Planck function at the electron temperature $T_e$, 
and $\alpha_\mathrm{B} = 2\times10^{-13}$ cm$^{-3}$ s$^{-1}$ is the case-B hydrogen recombination coefficient. 
The optical depth 
$\tau_\mathrm{ff}$ is almost a direct observable that depends only on the intensity and the assumed $T_e$. 
However, in equations (4) and (5) the EM, $Q_0$, and $n_e$ are not independent quantities. 
By definition, EM is integrated in the line of sight across the ionized object, thus under spherical geometry the diameter of the emitting \textsc{Hii} region is $\int dl = D$. In practice, EM is calculated first from $\tau_\mathrm{ff}$ using Eq. (4), and then $n_e$ and $Q_0$ are calculated using Eqs. (4) and (5).  
Our spherical approximation will be good for \textsc{Hii} regions that are isolated and round. For example, for the central \textsc{Hii} region in G010.6, from 3D radiative transfer modelling of the H$30\alpha$ line, the inferred $Q_0$ is 6 to $7\times10^{48}$ s$^{-1}$
\citep{GM23}, which is consistent the spherical value reported here: $5.1\pm1.4 \times10^{48}$ s$^{-1}$. 

Table \ref{tab:cont_params} lists the average physical parameters derived for the \textsc{HII} regions in the ALMA-IMF protoclusters, ordered by the sum of the $Q_0$ of the \textsc{Hii} regions in each protocluster. The spectral types of the ionizing stars are determined using the calibrations of \citet{Martins05} for O-type,  and of \citet{Panagia73} for B-type. A single ionizing star is assumed.   
For reference, Table \ref{tab:cont_params} also lists the evolutionary stage (young, intermediate, or evolved) of the ALMA-IMF protoclusters determined by \citet{Motte22} from a combination of mid- and far-IR emission, and also considering the H$41\alpha$ surface brightness in preliminary cubes. 
Our measurements of the ionized gas are overall consistent with this classification, both in terms of EM and $Q_0$. The earliest spectral type and amount of OB stars in each protocluster is also consistent with the proposed evolutionary picture: the youngest regions are devoid of massive stars, followed by regions with a few B-type stars, and then the most active protoclusters harbor from a few to several massive stars, including O-type.  
However, a few sources deserve a special mention. 
The ALMA-IMF mosaics of W43-MM2 also cover the region known as MM13 \citep{Motte2003}, which corresponds to \textsc{Hii} region 2 at the southern border of the field of view. Therefore, only \textsc{Hii} region 1 is considered for the total $Q_0$ of MM2. After this consideration, MM2 is the young protocluster with the lowest rate of ionizing photons detected in the sample. 
Also, G351.77 and G327.29 have very similar values for $Q_0 \approx 0.5\times10^{47}$ s$^{-1}$. In \citet{Motte22}, G327.29 (young) and G351.77 (intermediate) are at the limit between these two evolutionary cathegories, so their similar ionizing phothon rates are consistent with the proposed classification. 
W51-E is found to have the second largest $Q_0$ ($7.4\pm1.7\times10^{48}$ s$^{-1}$) among the ALMA-IMF protoclusters. Since free-free emission is produced by a collisional process, the number of ionizing photons necessary to generate the observed radiation depends on the density squared rather than on the total amount of ionized gas. Therefore, the prevalence of bright and very small ($\sim 10^{-2}$ pc) \textsc{Hii} regions in W51-E raises its rank among the evolved protoclusters when $Q_0$ is considered. The opposite situation occurs to G012.80, which has the second largest free-free flux after G333.60. However, this emission is distributed over a rather large area, therefore the required photoionization rate is the smallest among the evolved protoclusters. We note that the absolute evolutionary stage of a protocluster should not be only determined by its ionizing-photon rate, but also by its remaining gas reservoir and substructure \citep[e.g.,][]{Pouteau23,DiazGonzalez23}. 

\bigskip

\begin{table*}
	\centering
    \footnotesize
	\begin{tabular}{ccccccccc} 
        \hline 
        \hline
		Protocluster & $d$ & $S_\mathrm{ff,3mm}$ & EM &$n_e$ & $D$ & $Q_0$ & SpT & Evol \\
        \textsc{Hii} region
		& [kpc] & [mJy] & [$\times10^7$ pc cm$^{-6}$] & [$\times10^4$ cm$^{-3}$] & [pc] & [$10^{47}$ s$^{-1}$] & &  \\
        \hline 
        \hline 
        W43-MM1 & 5.5 & -- & -- & -- & -- & -- & -- & Y \\ \hline 
        
        G338.93 & 3.9 & -- & -- & -- & -- & -- & -- & Y \\ \hline 
        
        G328.25 & 2.5 & -- & -- & -- & -- & -- & -- & Y \\ \hline 
        
        W43-MM2$^\dagger$ & 5.5 & -- & -- & -- & -- & $0.0227\pm0.0081$ & -- & Y \\
        1 & $5.5$ & $1.38\pm0.30$ & $0.40\pm0.14$ & $1.72\pm0.31$ & $0.014$ & $0.0227\pm0.0081$ & B1 & -- \\
        2 & $5.5$ & $25.5\pm1.7$ & $0.61\pm0.18$ & $1.14\pm0.16$ & $0.047$ & $0.42\pm0.12$  & B0.5 & -- \\ 
        \hline 

        G337.92 & 2.7 & -- & -- & -- & -- &  $0.0363\pm0.0097$ & -- & Y \\
        1 & $2.7$ & $7.68\pm0.94$ & $0.144\pm0.045$ & $0.74\pm0.11$ & $0.026$ & $0.0304\pm0.0094$ & B1 & -- \\
        2 & $2.7$ & $1.48\pm0.41$ & $0.141\pm0.056$ & $1.10\pm0.22$ & $0.012$ & $0.0059\pm0.0023$ & B2 & -- \\
        \hline 

        G327.29 & 2.5 & -- & -- & -- & -- & $0.49\pm0.14$  & -- & Y \\
        1 & $2.5$ & $144.4\pm3.9$ & $0.199\pm0.057$ & $0.472\pm0.067$ & $0.089$ & $0.49\pm0.14$ & B0.5 & -- \\
        \hline

        G351.77 & 2.0 & -- & -- & -- & -- & $0.52\pm0.15$ & -- & I \\
        1 & $2.0$ & $230\pm17$ & $0.37\pm0.11$ & $0.75\pm0.11$ & $0.066$ & $0.50\pm0.15$ & B0.5 & -- \\
        2 & $2.0$ & $8.9\pm2.3$ & $0.135\pm0.051$ & $0.79\pm0.15$ & $0.022$ & $0.0194\pm0.0074$ & B1 & -- \\
        \hline 

        G353.41 & 2.0 & -- & -- & -- & -- & $1.54\pm0.33$ & -- & I \\
        1 & $2.0$ & $6.0\pm2.0$ & $0.123\pm0.055$ & $0.82\pm0.18$ & $0.018$ & $0.0130\pm0.0058$ & B1 & -- \\
        2 & $2.0$ & $240\pm13$ & $0.158\pm0.046$ & $0.391\pm0.056$ & $0.103$ & $0.52\pm0.15$ & B0.5 & -- \\
        3 & $2.0$ & $461\pm34$ & $0.81\pm0.24$ & $1.13\pm0.17$ & $0.063$ & $1.00\pm0.29$  & B0 & -- \\
        \hline

        G008.67 & 3.4 & -- & -- & -- & -- & $5.5\pm1.5$ & -- & I \\
        1 & $3.4$ & $834\pm32$ & $2.40\pm0.68$ & $1.69\pm0.24$ & $0.084$ & $5.2\pm1.5$ & O9.5 & -- \\
        2 & $3.4$ & $41.9\pm3.3$ & $0.338\pm0.099$ & $0.82\pm0.12$ & $0.050$ & $0.263\pm0.077$ & B0.5 & -- \\
        \hline 
        
        W43-MM3 & 5.5 & -- & -- & -- & -- & $12.0\pm3.4$ & -- & I \\
        1 & $5.5$ & $733\pm12$ & $0.99\pm0.28$ & $0.71\pm0.10$ & $0.199$ & $12.0\pm3.4$ & O8.5 & -- \\
        \hline 

        G012.80 & 2.4 & -- & -- & -- & -- & $46\pm12$ & -- & E \\
        1 & $2.4$ & $13490\pm230$ & $1.12\pm0.32$ & $0.567\pm0.080$ & $0.349$ & $42\pm12$ & O7 & -- \\
        2 & $2.4$ & $113\pm11$ & $0.221\pm0.066$ & $0.554\pm0.083$ & $0.072$ & $0.36\pm0.11$ & B0.5 & -- \\
        3 & $2.4$ & $1231\pm62$ & $0.90\pm0.26$ & $0.87\pm0.13$ & $0.118$ & $3.9\pm1.1$ & O9.5 & -- \\
        \hline 

        G010.62 & 5.0 & -- & -- & -- & -- & $57\pm14$ & -- & E \\
        1 & $5.0$ & $3823\pm64$ & $5.3\pm1.5$ & $1.73\pm0.25$ & $0.177$ & $51\pm14$ & O7 & -- \\
        2 & $5.0$ & $89.2\pm4.5$ & $0.94\pm0.27$ & $1.22\pm0.17$ & $0.064$ & $1.19\pm0.34$ & B0 & -- \\
        3 & $5.0$ & $344.4\pm9.8$ & $1.24\pm0.35$ & $1.06\pm0.15$ & $0.110$ & $4.6\pm1.3$ & O9.5 & -- \\
        \hline 

        W51-IRS2 & 5.4 & -- & -- & -- & -- & $63\pm18$ & -- & E  \\
        1 & $5.4$ & $3969\pm36$ & $2.53\pm0.72$ & $0.94\pm0.13$ & $0.284$ & $63\pm18$ & O6.5 & -- \\
        2 & $5.4$ & $1.85\pm0.49$ & $0.65\pm0.25$ & $2.32\pm0.45$ & $0.012$ & $0.029\pm0.011$ & B1 & -- \\
        \hline 

        W51-E & 5.4 & -- & -- & --- & -- & $74\pm17$ & -- & I \\
        1 & $5.4$ & $675\pm31$ & $13.3\pm3.8$ & $5.10\pm0.73$ & $0.051$ & $10.7\pm3.1$ & O8.5 & -- \\
        2 & $5.4$ & $106.4\pm5.0$ & $1.98\pm0.57$ & $1.94\pm0.28$ & $0.053$ & $1.69\pm0.48$  & B0 & -- \\
        3 & $5.4$ & $3717\pm27$ & $1.52\pm0.43$ & $0.653\pm0.092$ & $0.355$ & $59\pm17$ & O6.5 & -- \\
        4 & $5.4$ & $15.4\pm2.5$ & $3.7\pm1.2$ & $5.02\pm0.82$ & $0.015$ & $0.244\pm0.080$ & B0.5 & -- \\
        5 & $5.4$ & $182.6\pm6.3$ & $1.36\pm0.39$ & $1.28\pm0.18$ & $0.083$ & $2.89\pm0.82$ & O9.5 & -- \\
        \hline 

        G333.60 & 4.2 & -- & -- & -- & -- & $341\pm95$ & -- & E \\
        1 & $4.2$ & $10.0\pm1.6$ & $0.34\pm0.11$ & $1.06\pm0.17$ & $0.030$ & $0.095\pm0.031$ & B0.5 & -- \\
        2 & $4.2$ & $4.8\pm1.0$ & $0.33\pm0.12$ & $1.23\pm0.22$ & $0.021$ & $0.046\pm0.016$ & B1 & -- \\
        3 & $4.2$ & $71.2\pm4.5$ & $0.53\pm0.15$ & $0.91\pm0.13$ & $0.065$ & $0.68\pm0.20$ & B0 & -- \\
        4 & $4.2$ & $35200\pm190$ & $2.38\pm0.67$ & $0.592\pm0.084$ & $0.679$ & $338\pm95$ & O4 & -- \\
        5 & $4.2$ & $10.9\pm1.7$ & $0.33\pm0.10$ & $1.00\pm0.16$ & $0.032$ & $0.104\pm0.034$ & B0.5 & -- \\
        6 & $4.2$ & $78.5\pm4.5$ & $0.46\pm0.13$ & $0.79\pm0.11$ & $0.073$ & $0.75\pm0.22$ & B0 & -- \\
        7 & $4.2$ & $79.2\pm4.4$ & $0.39\pm0.11$ & $0.69\pm0.10$ & $0.080$ & $0.76\pm0.22$ & B0 & -- \\
        8 & $4.2$ & $67.2\pm4.1$ & $0.326\pm0.094$ & $0.637\pm0.092$ & $0.080$ & $0.64\pm0.19$ & B0  & -- \\
        \hline 
        \hline 
    \end{tabular}
    \caption{
    {\footnotesize
    \textsc{Hii} region parameters from free-free continuum estimations. Columns: (1) protocluster name; (2) distance in kpc, see \citet{Motte22}; (3) Free-free flux at 98.5 GHz; (4) emission measure; (5) average electron density; (6) source diameter; (7) hydrogen ionizing-photon rate; (8) spectral type from \citet{Martins05} for O-type stars, and from \citet{Panagia73} for B-type; (9) Evolutionary stage from \citet{Motte22}. $\dagger$ Only \textsc{Hii} region 1 in Fig. \ref{fig:ff-maps} corresponds to W43-MM2. \textsc{Hii} region 2 corresponds to the region known as MM13, south of MM2  \citep{Motte2003}. Error calculations include the assumed $20~\%$ error in $T_e$, which often dominates over the S/N in the photometry. Errors are reported to two significant figures. 
    }
    }
	\label{tab:cont_params}
\end{table*}   

\begin{figure*}
	\includegraphics[width=\textwidth]{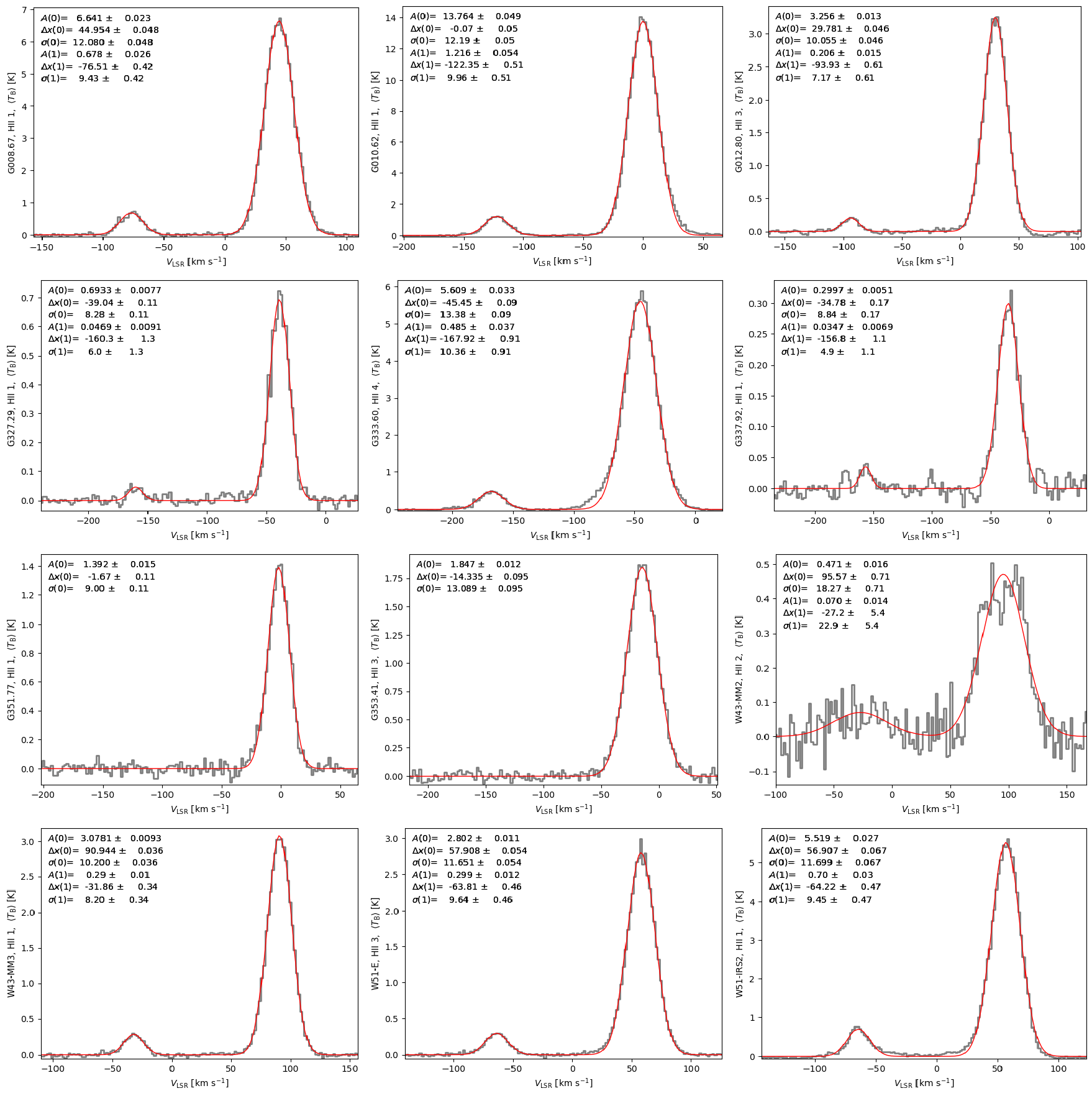}
    \caption{\footnotesize{Fits to H$41\alpha$ ($\nu_0 = 92.034434$ GHz) and He$41\alpha$ ($\nu_0 = 92.071938$ GHz) recombination line emission in the 12 protoclusters with detection.  This figure shows the brightest region for each protocluster. The rest of the fits are shown in Fig. \ref{fig:H_He_fitplots_app1} in the Appendix. The fitted parameters and derived abundances are listed in Table \ref{tab:H41_fits}.}
    }
    \label{fig:H_He_fitplots}
\end{figure*}

\begin{figure}
    \centering
	\includegraphics[width=0.6\columnwidth]{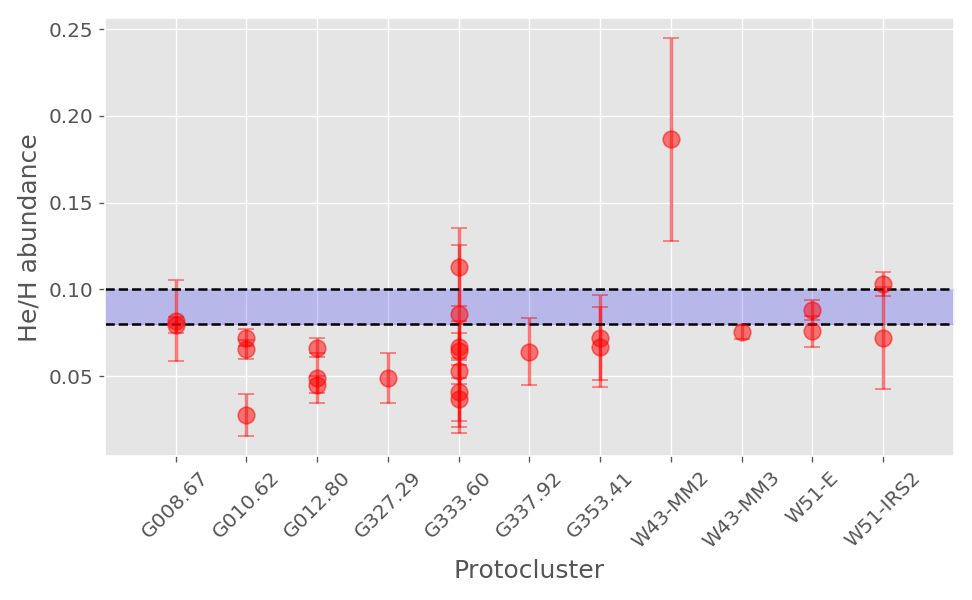}
    \caption{Number abundance of helium with respect to hydrogen $N_\mathrm{He} / N_\mathrm{H}$ for the \textsc{Hii} regions in the 11 ALMA-IMF protoclusters with detection H$41\alpha$ and He$41\alpha$. The shaded area within dashed lines mark the expected range for the Milky Way ISM, 0.08 to 0.1.}
    \label{fig:H_He_abund}
\end{figure}

\bigskip 
\section{Helium abundances} \label{sec:line_params}

Compared to optical and IR lines, radio recombination lines are free of obscuration and their excitation is simpler \citep{GS02}. 
As mentioned in Section \ref{sec:data}, the H$41\alpha$ cubes cover the frequencies of the neighboring He$41\alpha$ line.
Abundances of helium with respect to hydrogen can be derived under the assumption that for a given quantum number the recombination lines of the different atoms trace the same gas \citep[e.g.,][]{Wilson09}. The main caveat to this method is that the first ionization potentials  of these elements are $\chi = 13.6$ and 24.6 eV, 
respectively. Therefore, this assumption is valid only for massive stars. 
We calculate the number abundance of singly-ionized He to ionized H using the following equation \citep[for a review of the topic, see][]{RoelfsemaGoss92}: 
 
\begin{equation}
\frac{N_\mathrm{H^+}}{N_\mathrm{He^+}} = \frac{\int {\rm H}_{41\alpha} dv}{\int {\rm He}_{41\alpha} dv} \approx \frac{N_\mathrm{H}}{N_\mathrm{He}}. 
\end{equation}

\noindent
Given the very large potential for the second ionization of Helium  $\chi_\mathrm{He^+} = 54.40$ eV, this equation also assumes that in the \textsc{Hii} gas the number ratio of ionized hydrogen to singly-ionized helium is approximately equal to their total elemental abundance ratio.

For each protocluster, we calculated average abundances in the masks used to define the \textsc{Hii} regions shown in Fig. \ref{fig:ff-maps}, and with properties reported in Table \ref{tab:cont_params}.   
We performed Gaussian fitting using \texttt{Pyspeckit} \citep{Ginsburg22_pyspeckit}. The spectra were converted to units of brightness temperature as a function of LSR velocity prior to the fitting.
We use the H$41\alpha$ cubes without primary-beam correction to have a noise closer to spatially uniform. This means that the brightness scale of the fitted lines (see Fig. \ref{fig:H_He_fitplots}) is not corrected, but this does not affect the measured abundances, which depend on the line ratio within the mask. 
An initial round of fitting included only the hydrogen recombination line, and helium was added later using as initial guesses the parameters of the first fitting. For the relative He-to-H brightness, we took as initial guess a helium-to-hydrogen number abundance $N_\mathrm{He}/N_\mathrm{H} = 0.08$ (see Eq. \ref{eq:ff_map}), characteristic of the ISM in the Milky Way \citep{MendezDelgado20}. 

The intrinsically fainter He$41\alpha$ line at 92.071938 GHz can be more easily affected by molecular-line contamination. If present, the main contaminants  are lines of CH$_3$OCHO at 92.07402 GHz (ortho) and 92.073101 GHz (para). To avoid this contamination in the results, we rejected He$41\alpha$ fits with a peak brightness $> 0.3$ of the hydrogen peak. Additionally, we rejected results with a He$41\alpha$ centroid outside a range of $\pm10$ km s$^{-1}$ of its theoretical position at $-122.17$ km s$^{-1}$ from the H$41\alpha$. These criteria were found empirically (see Figs. \ref{fig:H_He_fitplots} and \ref{fig:H_He_fitplots_app1}). 
We also attempted fits with the C$41\alpha$ component. Valid fits including  C$41\alpha$ were found occasionally, but the extra fitted parameters had low S/N and did not improve the overall fit quality. The carbon line is expected to be fainter, and is only blueshifted by $-27.42$ km s$^{-1}$ with respect to helium. Also, its smaller ionization potential (11.3 eV) compared to hydrogen causes this line to be preferentially detected in photon-dominated regions \citep{Wyrowski97,Roshi05}, and not necessarily in \textsc{Hii} regions. 
The final fits only consider hydrogen and helium lines. These are shown in Figure \ref{fig:H_He_fitplots} for the \textsc{Hii} region in each of the 12 detected protoclusters with the highest average brightness. Fig. \ref{fig:H_He_fitplots_app1} in Appendix \ref{app:fitting} shows the rest of the  \textsc{Hii} regions. 

It is worth mentioning that although a Gaussian is a good model for most spectra, a few lines show excess or line wings. In the higher S/N spectra (peak $T_\mathrm{B} > 1$ K), G333.60 HII 4 (blueshifted) and  W51-E HII 1 (redshifted) show the clearest wings. G10.62 HII 1 (redshifted), G12.80 HII 1 (redshifted), G333.60 HII 6 (blueshifted), and W51-IRS2 HII 1 (blueshifted) also how milder wings. Apart from these, W43-MM2 HII 2 (MM13) is a double-peaked line. 

Figure \ref{fig:H_He_abund} shows the  helium-to-hydrogen number abundance ratio $N_\mathrm{He} / N_\mathrm{H}$ derived for the 25 \textsc{Hii} regions in the 11 protoclusters where $N_\mathrm{He} / N_\mathrm{H}$ was measured (see Table \ref{tab:H41_fits}).  
Note that the values that we report are relative number abundances. Appendix \ref{app:Y} clarifies the relation between our measurements and the often used relative He abundance per unit mass $Y$. 
A recent analysis using optical recombination lines of \textsc{Hii} regions in the Milky Way reports typical values of the quantity $12 + \log (N_\mathrm{He} / N_\mathrm{H})$ in the range from 10.9 to 11.0 \citep{MendezDelgado20}, or $N_\mathrm{He} / N_\mathrm{H} \approx$ 0.08 to 0.1. This range is broadly  consistent with our measurements in the UC \textsc{Hii} regions within the ALMA-IMF protoclusters. 

A few massive star formation regions have been reported to have number He abundances significantly offset from $0.1$, from $\sim~0.01$ to $\gtrsim 0.3$ \citep{RoelfsemaGoss92,DePree1997}. However, previous determinations have been usually done at cm-wavelengths with less sensitive observations\footnote{The brightness of radio and (sub)mm recombination lines scales almost linearly with frequency \citep[e.g.,][]{Peters12}.}.
Within our sample, there are 2 clear outliers with $N_\mathrm{He} / N_\mathrm{H}$ beyond $3\sigma$ of the 0.08 to 1 range, both of them on the lower abundance side: \textsc{Hii} region 2 in G010.62 with $0.028\pm0.012$, and \textsc{Hii} region 3 in G012.80 with $0.045\pm0.005$ (region 2 is at $2.2\sigma$). Their faint He$41\alpha$ emission compared to hydrogen can be seen in Fig. \ref{fig:H_He_fitplots_app1} for the former and Fig. \ref{fig:H_He_fitplots} for the latter. 
A proposed explanation for the apparent detection of small He abundances is a geometric effect, in which the He-ionized gas occupies a smaller volume than the H-ionized gas due to the larger ionization potential of helium   \citep{RoelfsemaGoss92,DePree1997}. This effect is less pronounced in resolved observations such as ours, but it could still be relevant. The derived spectral types for the ionizing stars of these \textsc{Hii} regions are O9.5 or later (see Table \ref{tab:cont_params}), therefore, it is plausible that a softer stellar spectrum in these regions is responsible for the comparatively smaller amounts of ionized helium. 

On the high end of the helium abundances, only \textsc{Hii} region 2 in W43-MM2 (also known as W43-MM13, see Section \ref{sec:hiis}) is  above $1\sigma$ from 0.1, with $N_\mathrm{He} / N_\mathrm{H} = 0.187\pm0.058$. 
We consider this region as a large He-abundance candidate, also considering its other special characteristics. The spectrum of this source also 
has the broadest H$41\alpha$  line among the sample ($43.0\pm1.7$ km s$^{-1}$, see Table \ref{tab:H41_fits}), and it is clearly double peaked (see Fig. \ref{fig:H_He_fitplots}). We propose that the ionized gas in this source could be produced by a stellar wind enriched by nuclear-processed products. 
It is known that massive stars can enrich the ISM with extra amounts of He and metals via powerful stellar winds \citep{Crowther07_review}, but it is unclear if these phenomena can exist at the $\sim$ ZAMS stages that we are probing. 
Finally, we note that \textsc{Hii} regions 1 and 2 in W51-E have a bright line at the expected position of He$41\alpha$ (see Fig. \ref{fig:H_He_fitplots_app1}). 
Our fitting procedure rejected it as due to molecular-line contamination, most likely from CH$_3$OCHO lines (see the previous explanation in this Section). 

\begin{table*}
	\centering
    \footnotesize
	\begin{tabular}{cccccccc} 
    \hline
    \hline
     & & H$41\alpha$ & & & He$41\alpha$ & & $N_\mathrm{He}/N_\mathrm{H}$ \\
    \hline 
    Protocluster & $A$ & $v_\mathrm{LSR}$ & FWHM & $A$ & $v_\mathrm{LSR}$ & FWHM & \\ 
    \textsc{Hii} region & [K] & [km s$^{-1}$] & [km s$^{-1}$] & [K] & [km s$^{-1}$] & [km s$^{-1}$] & \\    
    \hline 
    \hline
    G008.67 1 & $6.64\pm0.023$ & $44.954\pm0.048$ & $28.45\pm0.11$ & $0.678\pm0.026$ & $-76.51\pm0.42$ & $22.20\pm0.98$ & $0.0797\pm0.0047$ \\
    G008.67 2 & $0.930\pm0.016$ & $41.05\pm0.19$ & $22.52\pm0.44$ & $0.093\pm0.017$ & $-79.1\pm1.7$ & $18.4\pm3.9$ & $0.082\pm0.023$ \\
    \hline 
    G010.62 1 & $13.764\pm0.049$ & $-0.073\pm0.050$ & $28.69\pm0.12$ & $1.216\pm0.054$ & $-122.35\pm0.51$ & $23.4\pm1.2$ & $0.0722\pm0.0049$ \\
    G010.62 2 & $2.451\pm0.021$ & $5.916\pm0.089$ & $21.34\pm0.21$ & $0.076\pm0.022$ & $-117.5\pm2.7$ & $19.1\pm6.4$ & $0.028\pm0.012$ \\
    G010.62 3 & $4.451\pm0.017$ & $-0.874\pm0.041$ & $22.002\pm0.096$ & $0.337\pm0.018$ & $-123.38\pm0.50$ & $19.1\pm1.2$ & $0.0656\pm0.0054$ \\
    \hline 
    G012.80 1 & $3.162\pm0.012$ & $36.752\pm0.058$ & $30.48\pm0.14$ & $0.233\pm0.013$ & $-86.13\pm0.75$ & $27.6\pm1.8$ & $0.0665\pm0.0056$ \\
    G012.80 2 & $1.020\pm0.012$ & $36.44\pm0.12$ & $19.56\pm0.28$ & $0.084\pm0.016$ & $-85.7\pm1.1$ & $11.6\pm2.6$ & $0.049\pm0.014$ \\
    G012.80 3 & $3.256\pm0.013$ & $29.781\pm0.046$ & $23.68\pm0.11$ & $0.206\pm0.015$ & $-93.93\pm0.61$ & $16.9\pm1.4$ & $0.0452\pm0.0051$ \\
    \hline 
    G327.29 1 & $0.6933\pm0.0077$ & $-39.04\pm0.11$ & $19.50\pm0.25$ & $0.0469\pm0.0091$ & $-160.3\pm1.3$ & $14.1\pm3.1$ & $0.049\pm0.014$ \\
    \hline 
    G333.60 1 & $0.685\pm0.021$ & $-46.45\pm0.41$ & $27.40\pm0.96$ & -- & -- & -- & -- \\
    G333.60 2 & $0.593\pm0.015$ & $-47.09\pm0.29$ & $22.94\pm0.68$ & $0.050\pm0.015$ & $-169.0\pm3.5$ & $23.5\pm8.2$ & $0.086\pm0.040$ \\
    G333.60 3 & $1.961\pm0.022$ & $-44.77\pm0.11$ & $20.43\pm0.27$ & $0.158\pm0.025$ & $-163.9\pm1.3$ & $16.4\pm3.0$ & $0.065\pm0.016$ \\
    G333.60 4 & $5.609\pm0.033$ & $-45.446\pm0.090$ & $31.52\pm0.21$ & $0.485\pm0.037$ & $-167.92\pm0.91$ & $24.4\pm2.2$ & $0.0670\pm0.0078$ \\
    G333.60 5 & $0.5027\pm0.0092$ & $-33.13\pm0.18$ & $20.54\pm0.44$ & $0.041\pm0.013$ & $-154.6\pm1.6$ & $10.3\pm3.8$ & $0.041\pm0.020$ \\
    G333.60 6 & $0.953\pm0.017$ & $-50.49\pm0.22$ & $24.85\pm0.52$ & $0.072\pm0.025$ & $-173.4\pm2.1$ & $12.1\pm4.8$ & $0.037\pm0.019$ \\
    G333.60 7 & $0.913\pm0.017$ & $-45.54\pm0.20$ & $21.86\pm0.47$ & $0.048\pm0.017$ & $-166.9\pm3.9$ & $22.2\pm9.1$ & $0.053\pm0.029$ \\
    G333.60 8 & $0.679\pm0.010$ & $-44.61\pm0.17$ & $22.47\pm0.39$ & $0.082\pm0.011$ & $-166.1\pm1.3$ & $20.9\pm3.1$ & $0.113\pm0.022$ \\
    \hline 
    G337.92 1 & $0.2997\pm0.0051$ & $-34.78\pm0.17$ & $20.82\pm0.41$ & $0.0347\pm0.0069$ & $-156.8\pm1.1$ & $11.5\pm2.6$ & $0.064\pm0.019$ \\
    G337.92 2 & $0.2190\pm0.0063$ & $-32.74\pm0.30$ & $21.23\pm0.71$ & -- & -- & -- & -- \\
    \hline 
    G351.77 1 & $1.392\pm0.015$ & $-1.67\pm0.11$ & $21.19\pm0.27$ & -- & -- & -- & -- \\
    G351.77 2 & $0.561\pm0.083$ & $-9.8\pm1.2$ & $16.5\pm2.8$ & -- & -- & -- & -- \\
    \hline 
    G353.41 1 & $0.3239\pm0.0073$ & $-17.00\pm0.22$ & $20.38\pm0.53$ & $0.046\pm0.010$ & $-137.1\pm1.1$ & $10.3\pm2.6$ & $0.072\pm0.025$ \\
    G353.41 2 & $0.5077\pm0.0079$ & $-16.11\pm0.20$ & $25.50\pm0.46$ & $0.0366\pm0.0082$ & $-139.6\pm2.6$ & $23.7\pm6.2$ & $0.067\pm0.023$ \\
    G353.41 3 & $1.847\pm0.012$ & $-14.335\pm0.095$ & $30.82\pm0.22$ & -- & -- & -- & -- \\
    \hline 
    W43-MM2 1 & $0.96\pm0.17$ & $92.0\pm2.3$ & $26.6\pm5.5$ & -- & -- & -- & -- \\
    W43-MM2 2 & $0.471\pm0.016$ & $95.57\pm0.71$ & $43.0\pm1.7$ & $0.070\pm0.014$ & $-27.2\pm5.4$ & $54\pm13$ & $0.187\pm0.058$ \\
    \hline 
    W43-MM3 1 & $3.0781\pm0.0093$ & $90.944\pm0.036$ & $24.018\pm0.084$ & $0.289\pm0.010$ & $-31.86\pm0.34$ & $19.31\pm0.80$ & $0.0755\pm0.0041$ \\
    \hline 
    W51-E 1 & $16.81\pm0.13$ & $59.17\pm0.10$ & $27.13\pm0.25$ & -- & -- & -- & -- \\
    W51-E 2 & $3.794\pm0.052$ & $56.35\pm0.15$ & $21.87\pm0.34$ & -- & -- & -- & -- \\
    W51-E 3 & $2.802\pm0.011$ & $57.908\pm0.054$ & $27.44\pm0.13$ & $0.299\pm0.012$ & $-63.81\pm0.46$ & $22.7\pm1.1$ & $0.0882\pm0.0056$ \\
    W51-E 4 & $1.485\pm0.045$ & $51.79\pm0.42$ & $27.95\pm0.98$ & -- & -- & -- & -- \\
    W51-E 5 & $2.458\pm0.017$ & $67.307\pm0.076$ & $21.93\pm0.18$ & $0.273\pm0.021$ & $-53.68\pm0.56$ & $15.0\pm1.3$ & $0.0758\pm0.0089$ \\
    \hline 
    W51-IRS2 1 & $5.519\pm0.027$ & $56.907\pm0.067$ & $27.55\pm0.16$ & $0.704\pm0.030$ & $-64.22\pm0.47$ & $22.2\pm1.1$ & $0.1030\pm0.0069$ \\
    W51-IRS2 2 & $1.144\pm0.038$ & $48.90\pm0.46$ & $28.1\pm1.1$ & $0.256\pm0.067$ & $-71.5\pm1.2$ & $9.1\pm2.8$ & $0.072\pm0.029$ \\
    \hline 
    \hline
	\end{tabular}
 	\caption{{\footnotesize Fitted parameters for H$41\alpha$ and He$41\alpha$. Columns: (1) protocluster and \textsc{Hii} region; (2) and (5) peak intensity of spatially-averaged spectrum; (3) and (6) LSR velocity of line peak; (4) and (7) line $\mathrm{FWHM}$. Errors are reported to two significant figures.}}
	\label{tab:H41_fits}
\end{table*}

\section{Discussion} \label{sec:disc}

\subsection{On the subtraction of free-free contamination in (sub)mm observations of dust continuum}

Any millimeter survey aimed at studying dust continuum emission has the issue of potential free-free contamination. This becomes more relevant with the availability of sensitive observations of the innermost parts of star-forming cores and young stellar objects, where  free-free emission from ionized jets,  photoevaporating disks, and the birth of \textsc{Hii} regions are expected to contribute to the observed flux \citep[e.g.,][]{vdTM05,CG2012,Rota24}. 

There are three main observational methods to estimate the free-free contribution in millimeter maps: 
i) the most common method is using centimeter-wavelength images taken with, e.g., the VLA, and extrapolate them to the millimeter assuming a spectral index $\alpha_\mathrm{ff}$ for the free-free emission. This method usually assumes $\alpha_\mathrm{ff}=-0.1$, characteristic of the optically-thin regime (see Section \ref{sec:validation}); ii) creating spectral index $\alpha$ maps between two continuum bands, or measuring $\alpha$ within a given aperture. Free-free emission has a spectral index $\alpha_\mathrm{ff}$ that changes from 2 to $-0.1$ as frequency increases and the emission transitions from optically thick to thin. In contrast, the spectral index of dust emission in the radio and (sub)mm $\alpha_\mathrm{dust}$ has the opposite behavior with frequency, it changes from $\lesssim 4$ (optically thin) to 2 (thick) with increasing frequency. These opposite trends with frequency enable to create spectral index masks to exclude dust cores that are significantly contaminated by free-free \citep[e.g.,][]{Pouteau22}. \citet{DiazGonzalez23} provided spectral index maps sensitive to extended emission between 3 mm and 1 mm, using their combination of the ALMA-IMF continuum maps and single-dish surveys. This method, however, does not provide an estimate of the free-free emission in intensity units, but rather serve as rejection filter to separate contaminated from non-contaminated areas; and iii) the method that is systematically used in this paper. 

The main advantages of estimating the free-free contribution with our method are that all the surveyed protoclusters are observed with a $(u,v)$ coverage and beamsize close to the continuum images, and that the subsequent frequency scaling factor of the free-free estimation is more accurate and numerically closer to 1 (see Section \ref{sec:validation}).  

However, our proposed method has its own caveats, namely: 
i) the free-free map derived from the recombination line is less sensitive than the corresponding millimeter continuum image. Therefore, it is not able to detect the faintest ionized emission that could be present in the continuum maps; 
ii) the H$41\alpha$ emission is assumed to be in local thermodynamic equilibrium (LTE). This is a good assumption for most conditions; and 
iii) molecular line contamination of the recombination line emission can be an issue in a few very line-rich objects objects \citep[e.g.,][]{Law2021,Brouillet22}.

\subsection{A search for sites with excess recombination line emission} \label{sec:LTE-validity}

In this section, we look for regions within the ALMA-IMF images where the H$41\alpha$ line could be brighter than the LTE expectation. 
There is evidence for the rare occurrence of  (mostly weak) recombination line masers in massive star formation regions. 
The massive young star MonR2-IRS2 has a (sub)mm  maser with an amplification factor in the range 2 to 3 with respect to LTE \citep{JS20}. This type of maser is expected to occur in the (sub)mm range if the density conditions are appropriate \citep{Walmsley90,Peters12,Zhu22}.

We looked for areas in our maps which could be the sites of amplification of the H$41\alpha$ line.
For this, we inspected the pure dust estimation maps (total intensity minus free-free maps after convolution and regridding, see Section \ref{sec:proc}) and looked for areas that are significantly negative, i.e., that have negative ``potholes''. A naive interpretation of those potholes would be that they are regions where the H$41\alpha$ is brighter than the LTE assumption in Equation \ref{eq:ff_map}. 
We applied the following criteria for our search:  
i) that the pothole is clearly visible both in the Band 3 and Band 6 subtraction (pure dust) maps;   
ii) that the Band 3 and Band 6 potholes have an area larger than one beamsize (this is to avoid negative fluctuations of a few pixels in size); and 
iii) that the absolute value of the pothole minimum be larger than the respective error in the free-free estimation, both in Band 3 and Band 6. 

\begin{figure*}
	\includegraphics[width=\textwidth]{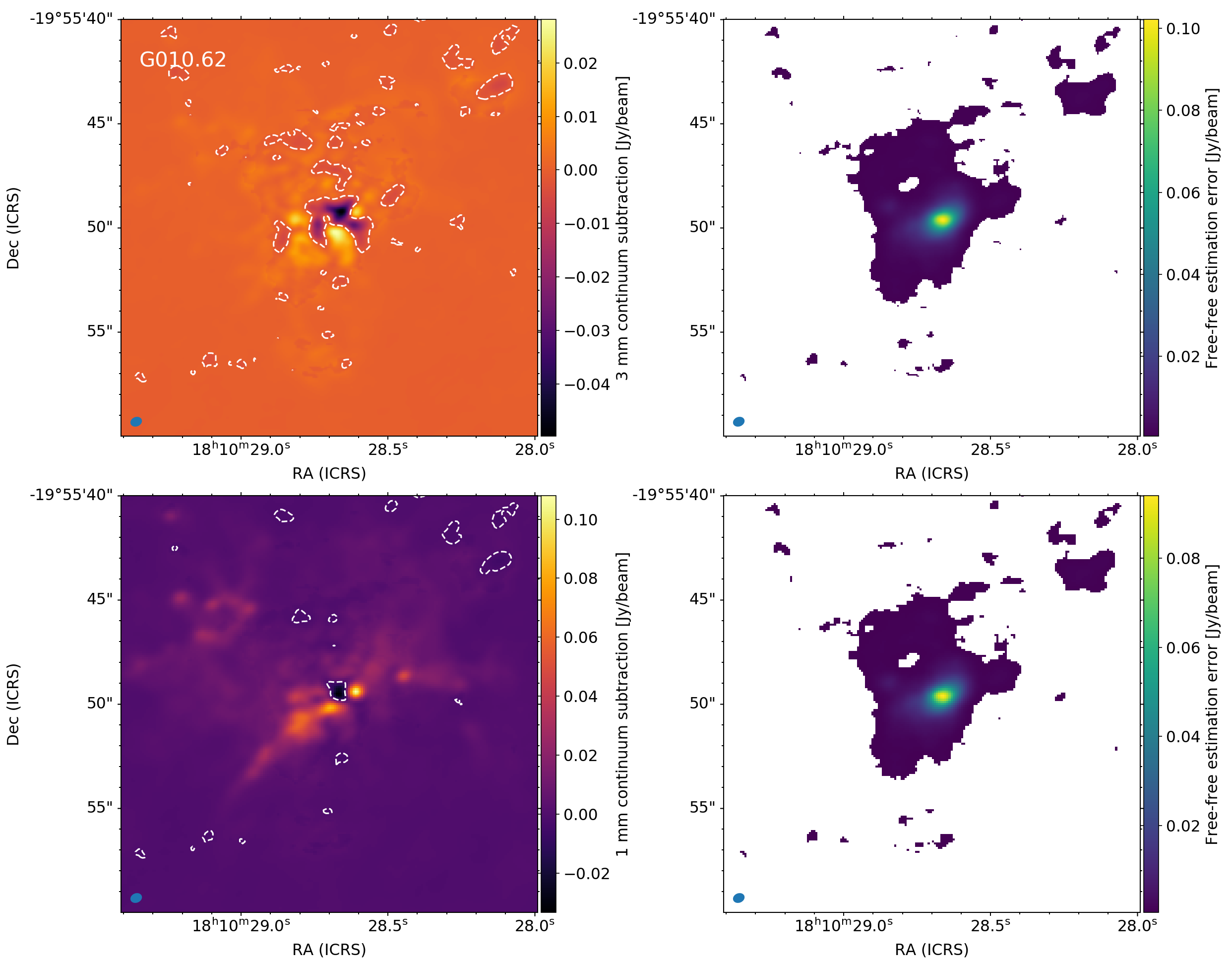}
    \caption{Detailed view of areas of the free-free oversubtraction issue, or negative ``potholes'' in G010.62.  
    Top panels: 3-mm continuum minus free-free estimation (left) and corresponding error of the free-free estimation (right). Dashed contours are at $-10\times \sigma_\mathrm{mad\_std} = - 8.2$ mJy beam$^{-1}$. 
    Bottom panels: B6 continuum minus free-free estimation (left) and corresponding error of the free-free estimation (right). Dashed contours are at $-10\times \sigma_\mathrm{mad\_std} = - 12.6$ mJy beam$^{-1}$. 
    ``Potholes'' in the central UC \textsc{Hii} region are present in both the 3 mm 1 mm subtractions, but the negative peak is still consistent with zero when the error map of the free-free estimation is considered.}
    \label{fig:potholes-maser}
\end{figure*}

We found that when the respective error maps are taken into account (third criterion above), all the cases potholes are consistent with zero. Therefore, we do not consider them of high significance. 
Fig. \ref{fig:potholes-maser} shows the case of  G010.62, where a clear ``pothole'' is seen at the position of the well-known UC \textsc{Hii} region at the center \citep[e.g.,][]{KetoWood2006,GM23}. This target was our clearest candidate, yet the central hole is within $1\sigma$ of the free-free estimation error at that position. 
We conclude that there is no significant evidence for H$41\alpha$ masers in the ALMA-IMF data, and that the areas where the continuum appears to be oversubtracted, notably in the 3 mm maps, are due to the inherent uncertainties in the derivation of the free-free estimations (see Eqs. \ref{eq:ff_map} and \ref{eq:err_ff}).

\subsection{Evolution of the \textsc{Hii} regions in the ALMA-IMF sample}

\begin{figure}
    \centering
	\includegraphics[width=0.42\columnwidth]{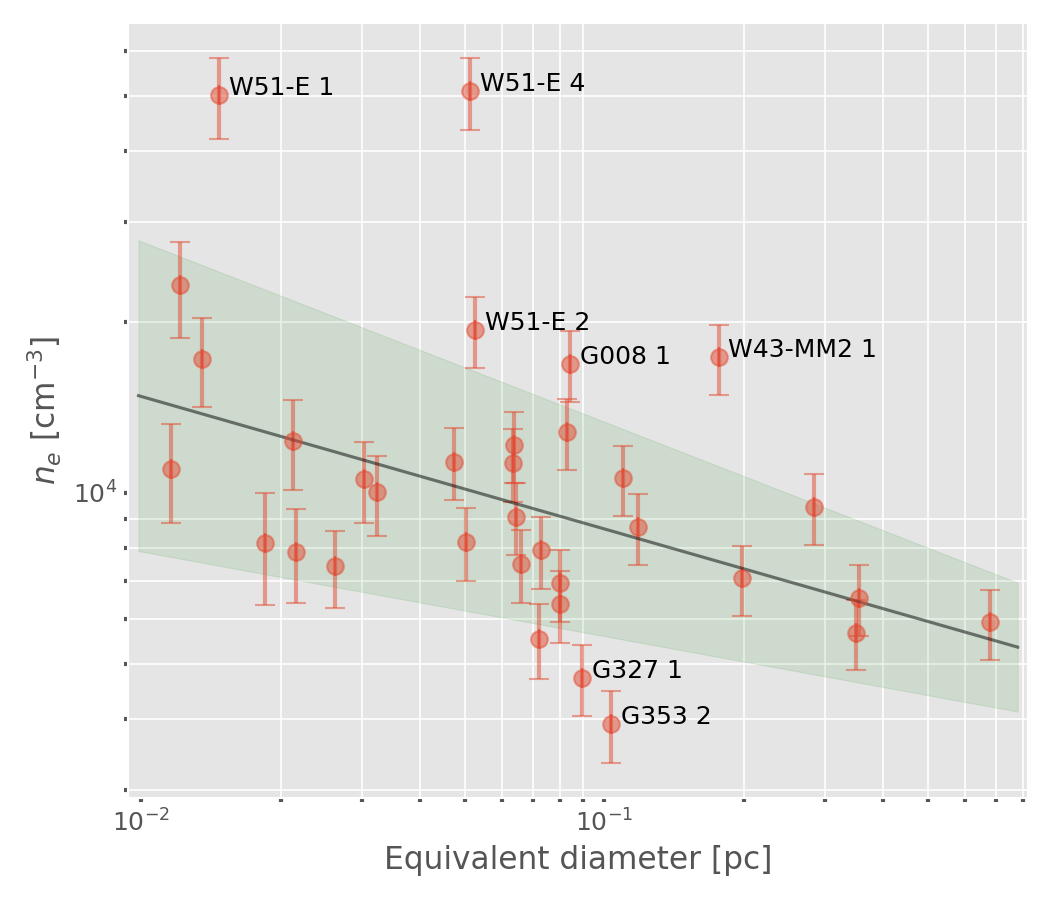} \\
 	\includegraphics[width=0.42\columnwidth]{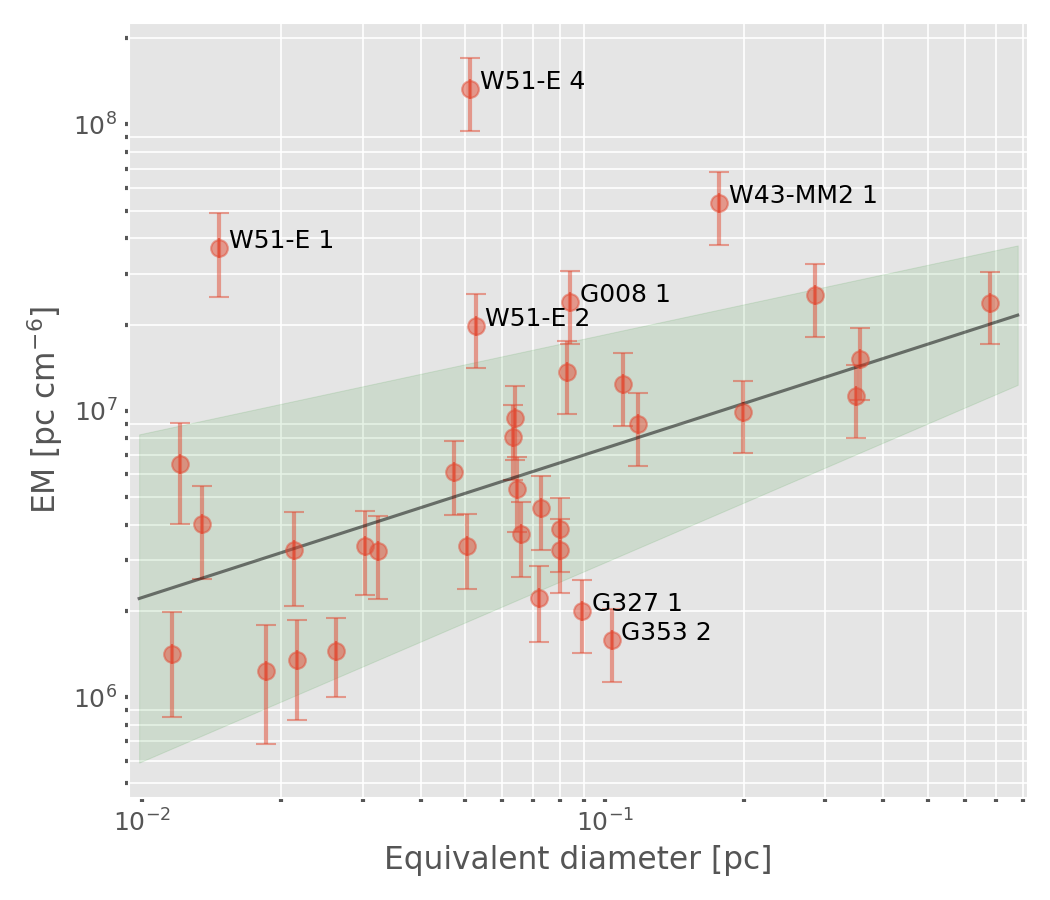} \\
  	\includegraphics[width=0.42\columnwidth]{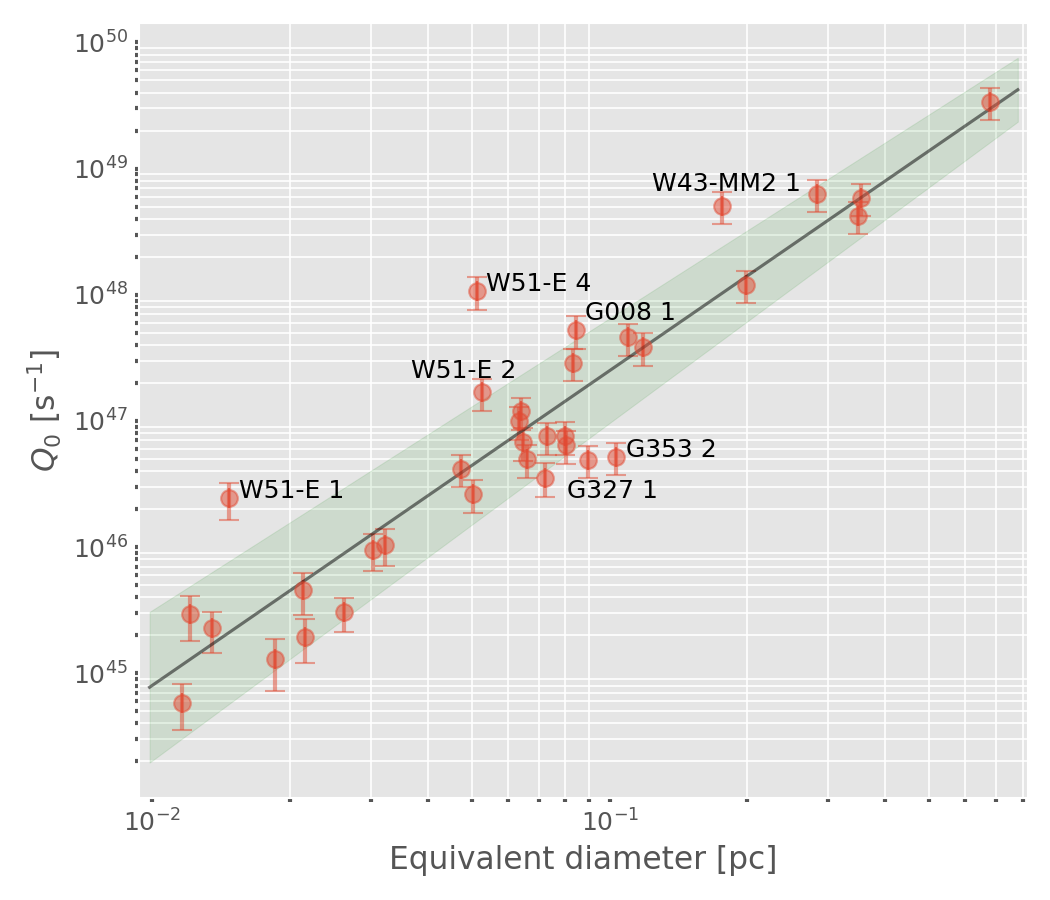} 
    \caption{Physical properties of the \textsc{Hii} regions in the ALMA-IMF protoclusters listed in Table \ref{tab:cont_params}, as a function of their equivalent diameter $D$. 
    \textit{Top} panel: electron density $n_e$,  \textit{middle} panel: emission measure EM,  \textit{bottom} panel: ionizing-photon rate $Q_0$. 
    The black lines and shaded areas show the respective best fits and $1\sigma$ confidence intervals. 
    The better constrained linear relation is  $\log [Q_0 / \mathrm{s}^{-1}] = (2.49\pm0.18) \times \log [D/\mathrm{pc}] + (49.89\pm0.23)$. 
    The slope and intercept for $\log (n_e)$ vs $\log (D)$ are $-0.23\pm0.08$ and $3.70\pm0.10$. 
    The slope and intercept for $\log (\mathrm{EM})$ vs $\log (D)$ are $0.52\pm0.17$ and $7.39\pm0.22$ ($1\sigma$ errors). 
    \textsc{Hii} regions which are an outlier in at least one of their properties are labeled.}
    \label{fig:HIIprops_2}
\end{figure}

The ALMA-IMF targets were selected to sample the evolutionary stages of massive star formation in clusters, from quiescent clumps that are faint in the infrared bands and lack ionized gas, to active clumps with prominent \textsc{Hii} regions \citep{Motte22}. Using H$41\alpha$ cubes, we calculated the basic physical properties of their \textsc{Hii} regions, such as their emission measure (EM), electron density ($n_e$), and hydrogen-ionizing photon rate ($Q_0$). 
For the \textsc{Hii} regions reported in Table \ref{tab:cont_params}, 
Figure \ref{fig:HIIprops_2} shows these properties as a function of the effective diameter $D$ of the emission.  
$Q_0$ is well correlated with $D$, with a Pearson correlation coefficient $r$ and p-value  $(r,p) = (0.89,1.6\times10^{-12})$. 
In contrast, the correlations between $n_e$ and EM with $D$ are not as clear, with $(r,p) = (-0.25,0.15)$ and $(0.12,0.50)$, respectively.
A weighted least-squares fit to the $Q_0$ vs $D$ relation gives $\log [Q_0 / \mathrm{s}^{-1}] = (2.49\pm0.18) \times \log [D/\mathrm{pc}] + (49.89\pm0.23)$. 

From Eq. (5), the observed $Q_0 \propto D^{2.5}$ relation falls in between the expected relations if $n_e$ or EM were constant among the sample of \textsc{Hii} regions ($Q_0 \propto D^3$ or $Q_0 \propto D^2$, respectively). Motivated by this, we fit a posteriori the $n_e$ vs $D$ and EM vs $D$ relations in logarithmic space and find that their respective slope and intercept, with $1\sigma$ errors, are ($-0.23\pm0.08$, $3.70\pm0.10$) and ($0.52\pm0.17$, $7.39\pm0.22$). 
Taking $Q_0 \propto D^{2.5}$ as valid, it is indeed expected that $n_e \propto D^{-0.25}$ and $\mathrm{EM} \propto D^{0.5}$. 

Previous authors have reported inverse correlations between $n_e$ and $D$ in samples of UC \textsc{Hii} regions, with power-law slopes $\approx -1$ \citep[e.g.,][]{GarayLizano99,KimKoo01}. If the ionizing stars had constant $Q_0$ across the sample, then $n_e \propto D^{-1.5}$ is expected. However, there is no reason to think that all UC \textsc{Hii} regions will be ionized by stars of the same mass at the moment of observation, and to follow the same path on the $n_e$ vs $D$ diagram as they expand and rarefy. 
A possible cause for our differing results with respect to previous surveys is that those have focused on the brightest UC \textsc{Hii} regions within  rather evolved targets, selected from their mid-IR colors \citep[e.g.,][]{WC89,Kurtz94}. In contrast, the UC \textsc{Hii} regions in our sample are distributed in a broad range of protocluster evolutionary stages, and in different locations within a given protocluster. A homogeneous analysis of a larger sample could help to understand this issue. As pointed out by \citet{GarayLizano99}, a positive correlation in the $Q_0$ vs $D$ diagram (bottom panel of Fig. \ref{fig:HIIprops_2}) indicates that the smallest UC \textsc{Hii} regions are on average ionized by less massive stars. 
Therefore, it could be that the smaller \textsc{Hii} regions in our sample are not necessarily dynamically younger, or at least that the effect of the ionizing spectral type is larger than the effect of age. 
A lack of a direct relation between UC \textsc{Hii} region size and other measured properties has been previously proposed as a solution to the apparent (excess) lifetime ``problem'' of UC \textsc{Hii} regions \citep{Peters2010a,Peters2010b,DePree2014}. 
Another possible explanation for differing results in the $n_e$ vs $D$ relation is environment. For example, the recent study by \citet{Meng22} of the UC \textsc{Hii} regions in the extreme protocluster Sgr B2 found that $n_e$ and $D$ follow an inverse relation, but with significant scatter and at higher densities than previous studies in other regions. 
Similarly, \citet{Yang21} analyzed several samples of UC \textsc{HII} regions and found that their physical properties probably shift with evolution. 
Finally, the presence or absence of ionized haloes around UC \textsc{Hii} regions and the ability of interferometric observations to recover them might have an effect too \citep{KimKoo01}. 
Fig. \ref{fig:HIIprops_2} labels the outliers from the fitted relations. The most notable ones with an excess in $n_e$, EM, or $Q_0$ are \textsc{Hii} regions 1 and 4 in W51-E, followed by \textsc{Hii} region 1 in W43-MM2, 1 in G008.67, and 2 in W51-E. The UC \textsc{Hii} regions in the north-south dust ridge of W51-E (which we label 1 and 2) are known to have active infall signatures \citep[e.g.,][]{Zhang98,Goddi16}.

\begin{figure}
\centering 
	\includegraphics[width=0.45\columnwidth]{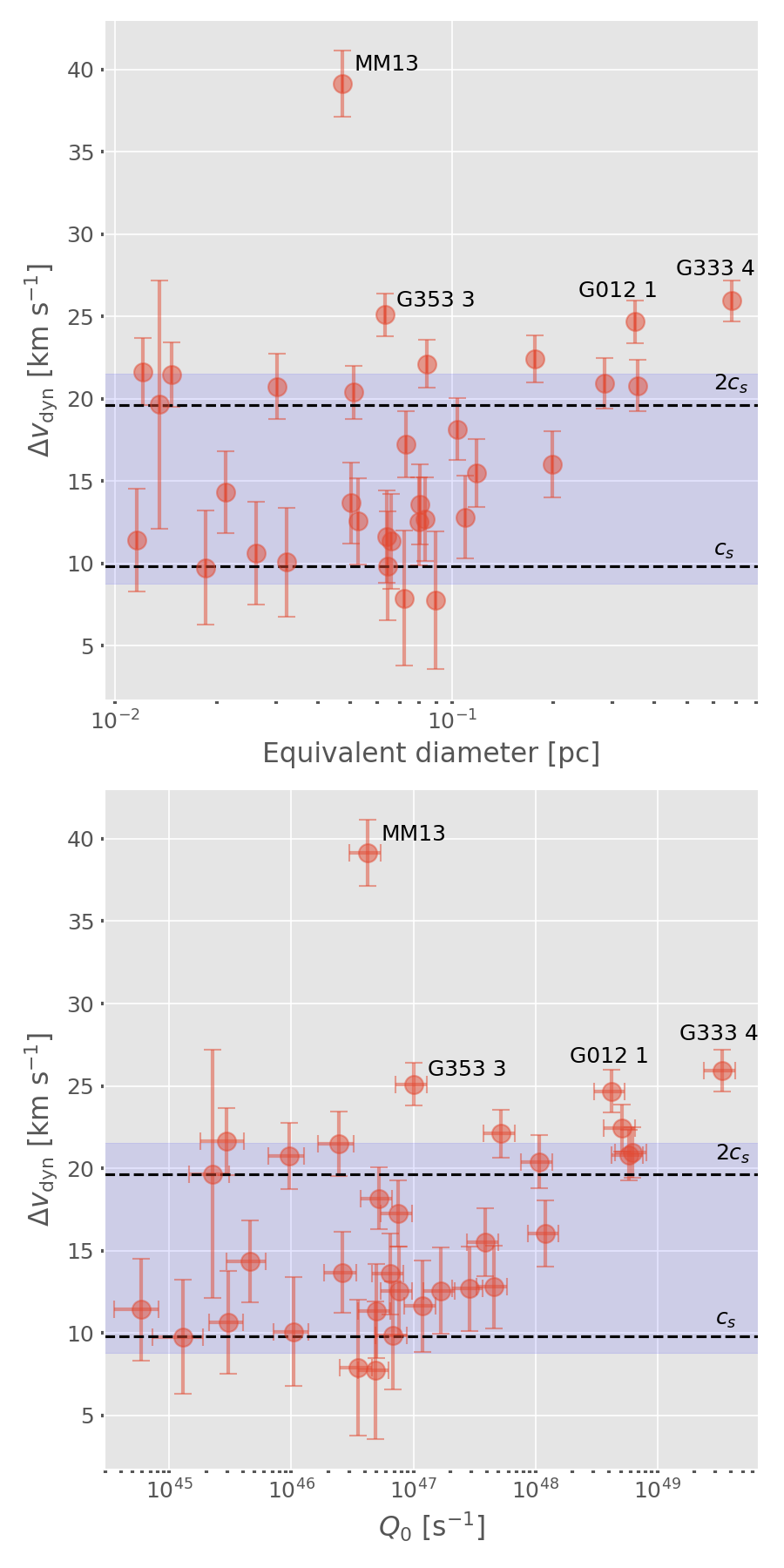}
    \caption{
    Dynamical FWHM linewidth ($\Delta V_\mathrm{dyn}$) of the H$41\alpha$ emission vs the logarithm of their equivalent diameter (top panel) and of the  ionizing-photon rate (bottom panel). The plot shows the \textsc{Hii} regions in the ALMA-IMF protoclusters listed in Table \ref{tab:cont_params}, except for \textsc{Hii} region 2 in G351.77, whose total linewidth is consistent with being purely thermal. The speed of sound of ionized gas at the assumed $T_e = 7000$ K is $c_s = 9.82$ km s$^{-1}$. Considering a $20~\%$ error in $Te$, the shaded area marks the expected range of \textsc{Hii} region expansion velocities between $c_s(T_e-\sigma_{T_e})$ and $2c_s(T_e+\sigma_{T_e})$. Outliers are marked.} 
    \label{fig:linewQ0}
\end{figure}

We now calculate the corresponding H$41\alpha$ dynamical linewidths $\Delta V_\mathrm{dyn}$. 
The total recombination linewidth is a non-linear sum of collissional (``pressure''), thermal, and dynamical widths \citep{GS02,KZK08}. Following the equations in the appendix of \citet{GM12}, we estimate that collisional broadening is about $0.05$ km s$^{-1}$ for densities as high as $10^5$ cm$^{-3}$, therefore it can be neglected. 
The thermal FWHM linewidth at $T_e=7000$ K $\pm 20~\%$ is $17.9\pm1.8$ km s$^{-1}$.
Supersonic recombination lines have been suggested to be a signpost of ultracompact and ``hypercompact'' \textsc{HII} regions with additional physical ingredients beyond the classical hydrodynamical expansion \citep[e.g.,][]{DePree04,GM09,Sandell09}. However, these objects appear to be very rare, including in our survey. 
Interestingly, $\Delta V_\mathrm{dyn}$ appears to be weakly correlated both with \textsc{Hii} region size (Fig. \ref{fig:linewQ0}, top panel) and ionizing-photon rate $Q_0$ (bottom panel), with $(r,p) = (0.34, 0.05)$ and $(0.33,0.06)$, respectively. 
As shown before, \textsc{Hii} region size and $Q_0$ are well correlated, therefore the trend of increasing dynamical linewidth with size likely reflects the same underlying physical process. 
The hydrodynamical expansion velocity of an \textsc{Hii} region is expected to be $\lesssim c_s$ \citep{Spitzer78,Bisbas15}, where the speed of sound of ionized gas is $c_s \approx 10$ km s$^{-1}$. Therefore, an optically-thin \textsc{Hii} region is expected to have an observed dynamical linewidth $\lesssim 2c_s$. Most of the objects in our sample satify this, as shown in Fig.  \ref{fig:linewQ0}. However, the trend of increasing $\Delta V_\mathrm{dyn}$ with $Q_0$ (spectral type), as well as the small excess linewidth above $2c_s$ at high $Q_0$, suggest that other acceleration mechanisms could play a role at high luminosities. 
Photoevaporative flows in density gradients, such as those in structured clouds, 
are able to accelerate \textsc{Hii} gas to a few tens of km s$^{-1}$ \citep{ArthurHoare2006,Zamora19}. Faster velocities would require the presence of stellar winds. We suggest that this is the case for the \textsc{Hii} region south of W43-MM2 (MM13).  This \textsc{Hii} region stands out with  $\Delta V_\mathrm{dyn} = 39.1\pm2.0$ km s$^{-1}$. An embedded stellar or disk wind \citep[e.g.,][]{Guzman20,GM23} could produce the observed large linewidth in this region, as well as in the other outlier objects with $\Delta V_\mathrm{dyn} > 2c_s$, which are labeled in Fig. \ref{fig:linewQ0}.  
Finally, we keep in mind that our analysis of helium abundances (see Section \ref{sec:line_params}) revealed that the \textsc{Hii} region MM13 also has the largest helium abundance $N_\mathrm{He}/N_\mathrm{H} = 0.187\pm0.058$, and that the line in this source is the only one which is clearly double-peaked. 
A full investigation of the ionizing sources requires the observation and modelling of more recombination lines.

\section{Conclusions} \label{sec:conclusions}

We made tailored H$41\alpha$ cubes using the ALMA-IMF pipeline \citep{Ginsburg22, Cunningham2023} to create  template maps, and their associated errors, of the free-free emission in the 3 mm and 1.3 mm continuum maps. With this paper we distribute the input cubes, the derived data products, and associated scripts, which can be used  to separate the free-free ``contamination'' from dust emission  in ALMA continuum maps for other projects. 

We validated our procedure against the more common approach of extrapolating the free-free emission from centimeter maps to  millimeter wavelengths. For our benchmark region, we found that both approaches give the same result within the errors for the vast majority of the pixels with emission in both tracers, but that they do not necessarily match in a few UC  \textsc{Hii} regions where either the recombination line can have excess brightness, or the optically thin ($\propto \nu^{-1}$) extrapolation from the centimeter to the millimeter is underestimation. Another possible issue  is molecular line contamination, which we found to be important only in three hot molecular cores. 

We used the derived free-free maps to estimate the properties of \textsc{Hii} regions across the ALMA-IMF protoclusters. We defined 34 regions of \textsc{Hii} emission and measured their photometry. Then, we derived their physical properties such as emission measure (EM), electron density ($n_e$), hydrogen ionizing-photon rate ($Q_0$), and spectral type (SpT) of the (single) ionizing star. 
$Q_0$ and SpT follow a trend consistent with the evolutionary picture proposed by \citet{Motte22}. The youngest protoclusters (W43-MM1, G338.93, G328.25) lack detectable emission from ionized gas. W43-MM2 and G337.92 harbor B-type stars with  $Q_0$ of few $\times10^{45}$ s$^{-1}$ (the MM13 region south of MM2 also contains a more luminous B-type star). G351.77 abd G327.29 contain early B-type stars with a total $Q_0$ of few $\times10^{46}$ s$^{-1}$. G353.41 and G008.67 have early B to late O-type stars with $Q_0\sim10^{47}$ s$^{-1}$. The following ALMA-IMF protoclusters (W43-MM3, G012.80, G010.62, W51-IRS2, and W51-E) contain from a few to several O and B-type stars, with $Q_0$ rising from $\sim 1\times$ to $7\times10^{48}$ s$^{-1}$. Finally, G333.60 contains the brightest and largest \textsc{Hii} region in the sample, as well as the largest number of OB stars, totalling $Q_0\approx3\times10^{49}$ s$^{-1}$. 

We performed measurements of the number abundance of He$^+$ with respect to H$^+$, which is a proxy to the total He-to-H abundance $N_\mathrm{He}/N_\mathrm{H}$ in \textsc{Hii} regions. Most of the 25 regions with He$^+$ measurement (see Table \ref{tab:H41_fits}) have values consistent with the expected  $N_\mathrm{He}/N_\mathrm{H} \approx 0.08$ in the Galactic ISM. Two \textsc{Hii} regions appear to have helium abundances  significantly below this value. We propose that this is due the relatively late spectral type of their ionizing stars, which produces He$^+$ regions which are smaller than their  respective H$^+$ regions. Only one  \textsc{Hii} region (W43 MM13, south of MM2) appears to have $N_\mathrm{He}/N_\mathrm{H}$ above the standard value.   

We looked for sites of significant amplification of the H$41\alpha$ with respect to LTE. This search returned negative results when the errors of the free-free estimation maps are considered. 

We investigated the possible correlations between the measured \textsc{Hii} region properties in the sample. 
The correlations $n_e$ and equivalent diameter $D$ and between EM and $D$ are weak. 
In contrast, $Q_0$ and $D$ are well correlated and have a linear dependence in log-log space with an exponent $\approx 2.5$.  
This favors an interpretation where the smaller UC \textsc{Hii} regions are not necessarily the less dynamically evolved versions of the larger ones, but that rather they are on average ionized by less massive stars. 
Finally, the dynamical width $\Delta V_\mathrm{dyn}$ of the \textsc{Hii} regions in the sample was deconvolved from the thermal and collisional widths, the latter of which is totally negligible. Moderate positive correlations were found between $\Delta V_\mathrm{dyn}$ and $D$, and between $\Delta V_\mathrm{dyn}$ and  $Q_0$. Dynamical widths vary from approximately the speed of sound of ionized gas $c_s$ to $\gtrsim 2c_s$, with MM13 being again an outlier with $\Delta V_\mathrm{dyn} \approx 4c_s$. The trend among the sample of increasing $\Delta V_\mathrm{dyn}$ with $Q_0$ suggests that, in the \textsc{Hii} regions ionized by the most massive stars, further mechanisms such as disk or stellar winds could play a role in their kinematics. 

\begin{acknowledgements}
We thank the anonymous referee for a detailed report, which helped to many aspects of this paper. 
R.G.M., D.D.G., and R.R.S. acknowledge support from UNAM-PAPIIT project IN108822 and from CONACyT Ciencia de Frontera project ID 86372. 
A.G. acknowledges support from the NSF under grants AST 2008101 and CAREER 2142300. 
T.Cs. has received financial support from the French State in the framework of the IdEx Université de Bordeaux Investments for the future Program. 
A.M.S. gratefully acknowledges support by the Fondecyt Regular (project code 1220610), and ANID BASAL project FB210003.
R.A.G. gratefully acknowledges support from ANID Beca Doctorado Nacional 21200897.
L.B. gratefully acknowledges support by the ANID BASAL project FB210003. 
\end{acknowledgements}

\software{
\texttt{Astropy} \citep{Astropy2013,Astropy2018}, \texttt{PySpecKit} \citep{Ginsburg22_pyspeckit}, \texttt{CASA} \citep{CASA2022}, APLpy \citep{APLpy}
}



\appendix

\section{Comparison of original, free-free, and pure-dust maps} \label{app:pure-dust}

Fig. \ref{fig:comparison-maps-app} shows comparison plots, covering the areas with free-free emission, of the original 1.3 mm continuum maps and the corresponding free-free and pure-dust estimations. 

\begin{figure*}
\centering
\includegraphics[width=1.00\linewidth]{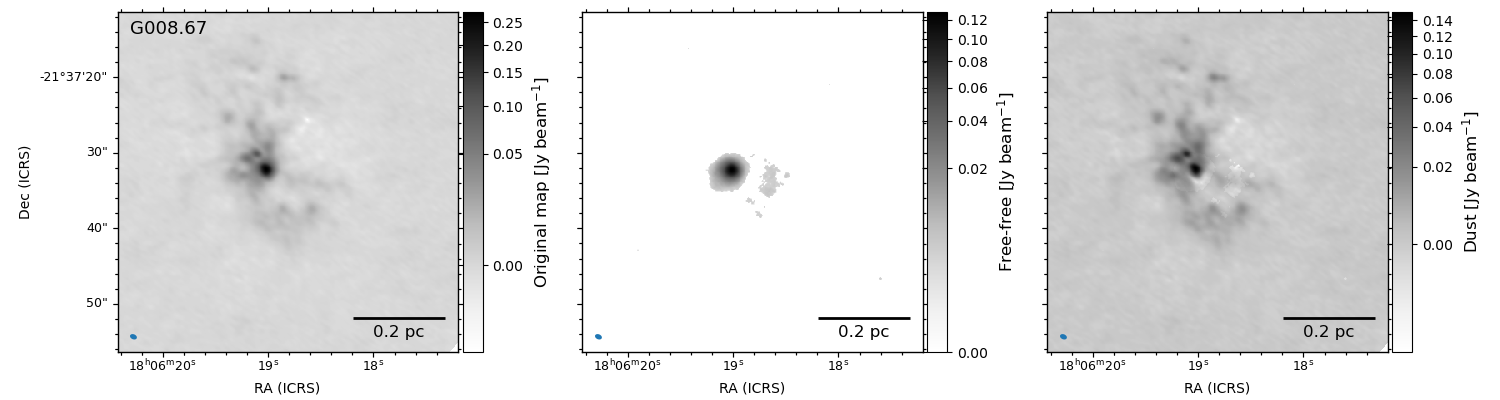}~\\
\includegraphics[width=1.00\linewidth]{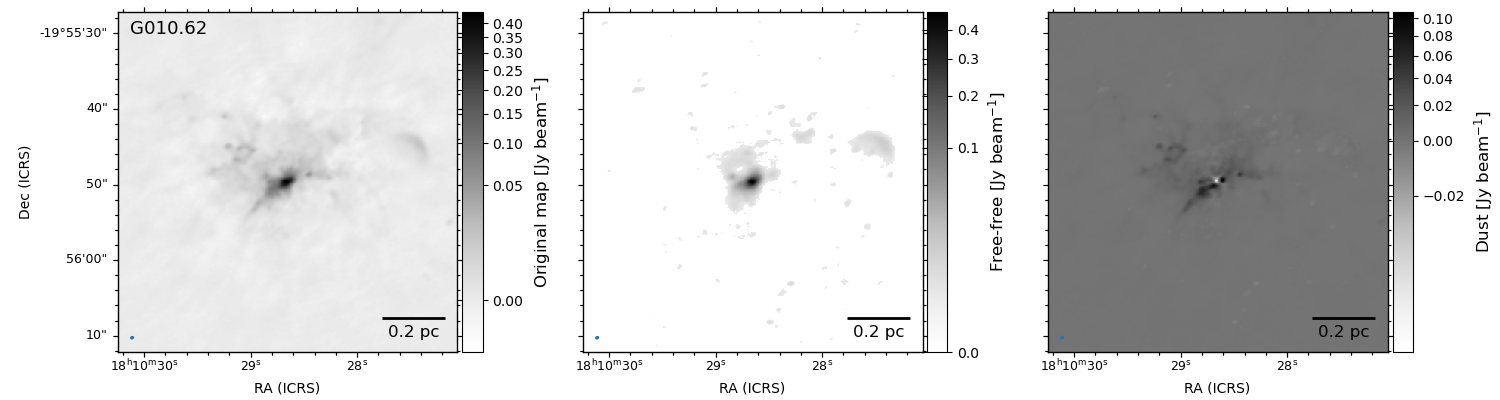}~\\
\includegraphics[width=1.00\linewidth]{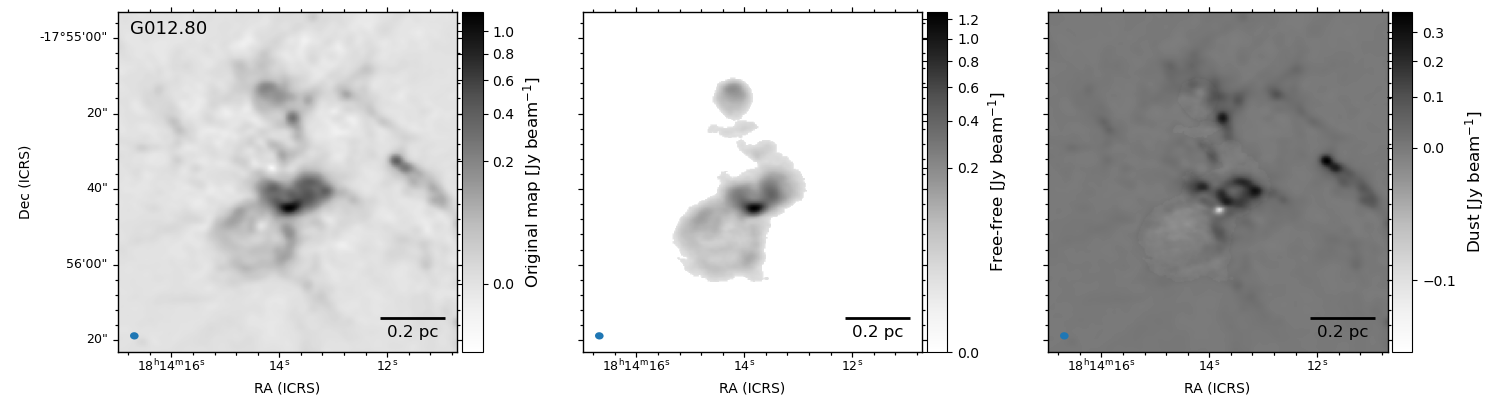}~\\
\includegraphics[width=1.00\linewidth]{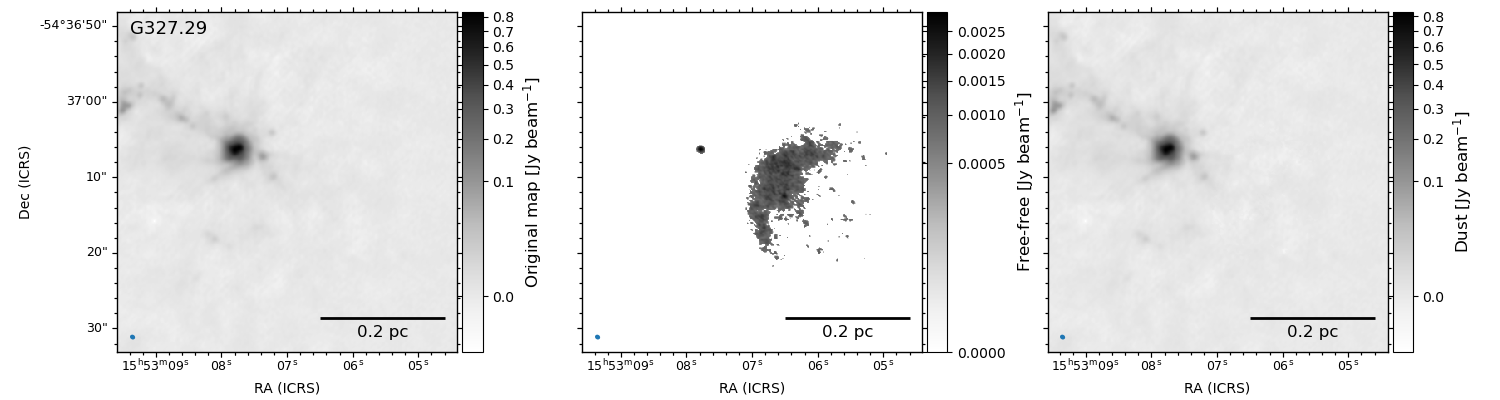}~
\caption{
Comparison of the original 1.3 mm continuum image convolved to the angular resolution of the H$41\alpha$ cube ({\it left} panel), free-free estimation map at 1.3 mm ({\it center}), and the corresponding pure-dust estimation map ({\it left}). The images shown are prior to primary-beam correction. The color normalization is a power law with a 0.33 exponent. The displayed areas are the same as in Fig. \ref{fig:ff-maps}, and correspond to where free-free emission was detected.    
}
\label{fig:comparison-maps-app}
\end{figure*}

\addtocounter{figure}{-1}
\begin{figure*}
\centering
\includegraphics[width=1.00\linewidth]{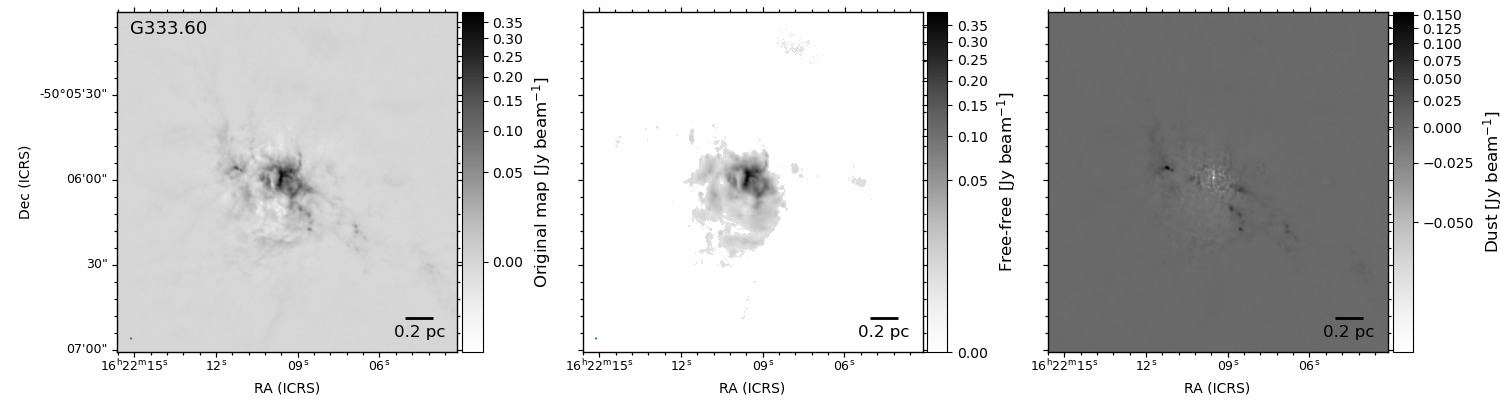}~\\
\includegraphics[width=1.00\linewidth]{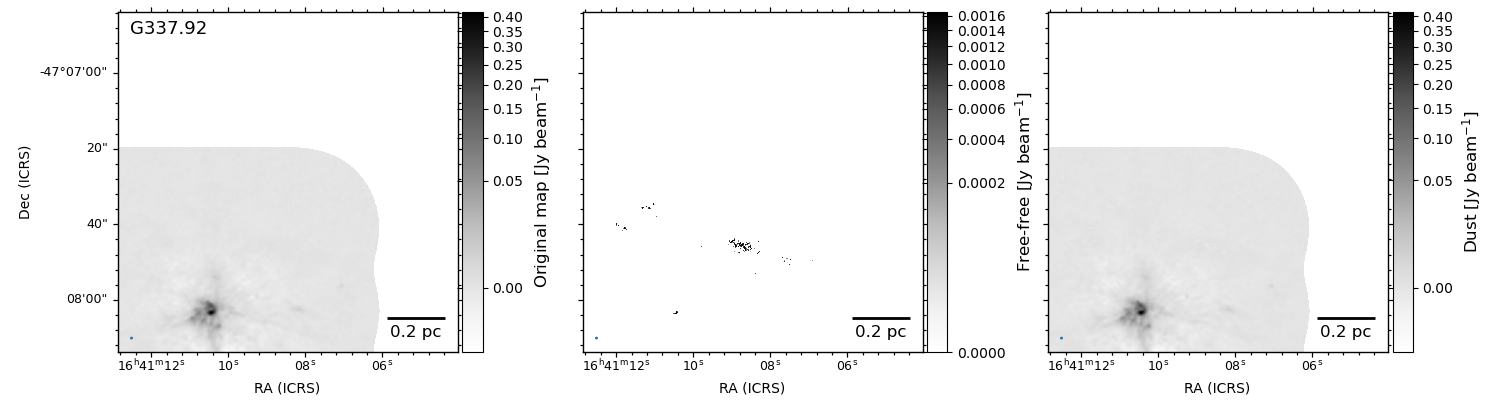}~\\
\includegraphics[width=1.00\linewidth]{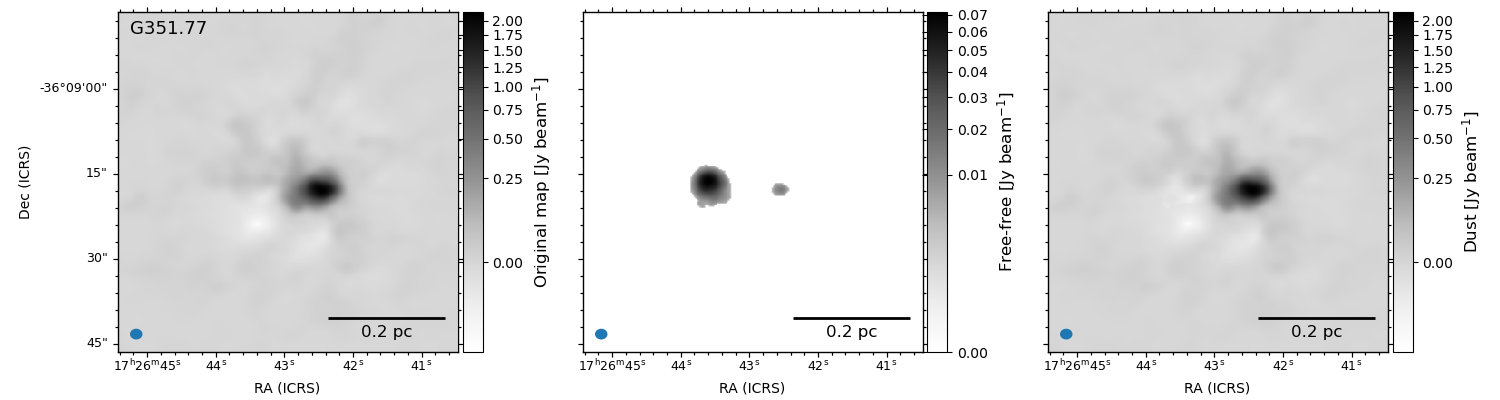}~\\
\includegraphics[width=1.00\linewidth]{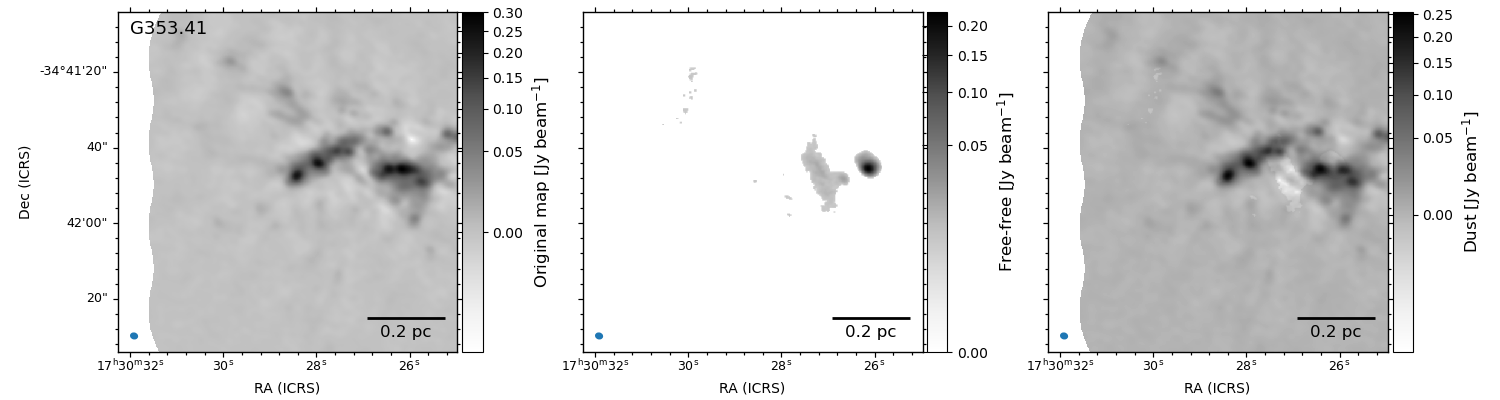}~
\caption{
Continued. 
Comparison of the original 1.3 mm continuum image convolved to the angular resolution of the H$41\alpha$ cube ({\it left} panel), free-free estimation map at 1.3 mm ({\it center}), and the corresponding pure-dust estimation map ({\it left}). The images shown are prior to primary-beam correction. The color normalization is a power law with a 0.33 exponent. The displayed areas are the same as in Fig. \ref{fig:ff-maps}, and correspond to where free-free emission was detected.    
}
\end{figure*}

\addtocounter{figure}{-1}
\begin{figure*}
\centering
\includegraphics[width=1.00\linewidth]{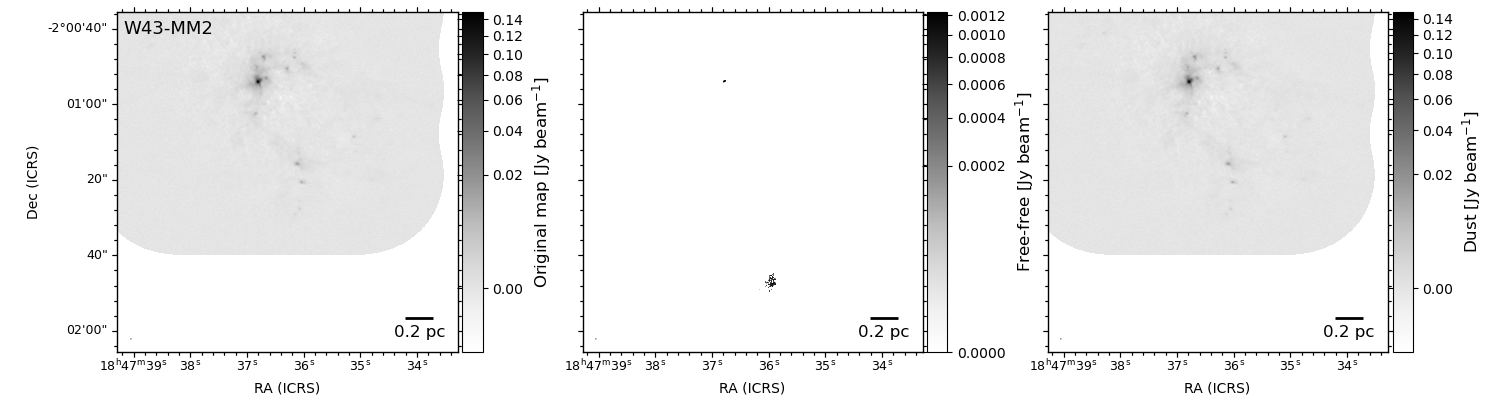}~\\
\includegraphics[width=1.00\linewidth]{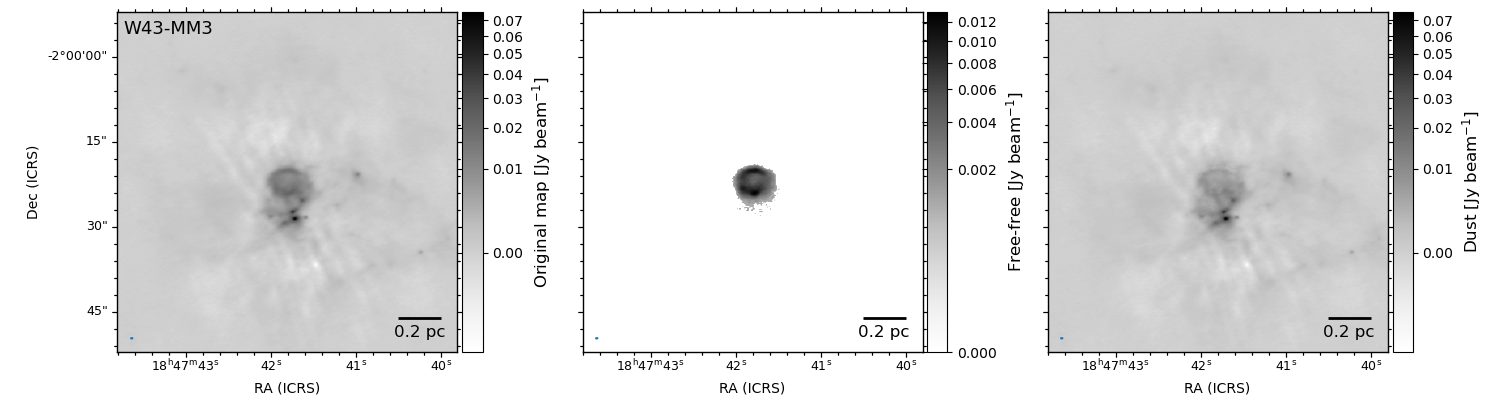}~\\
\includegraphics[width=1.00\linewidth]{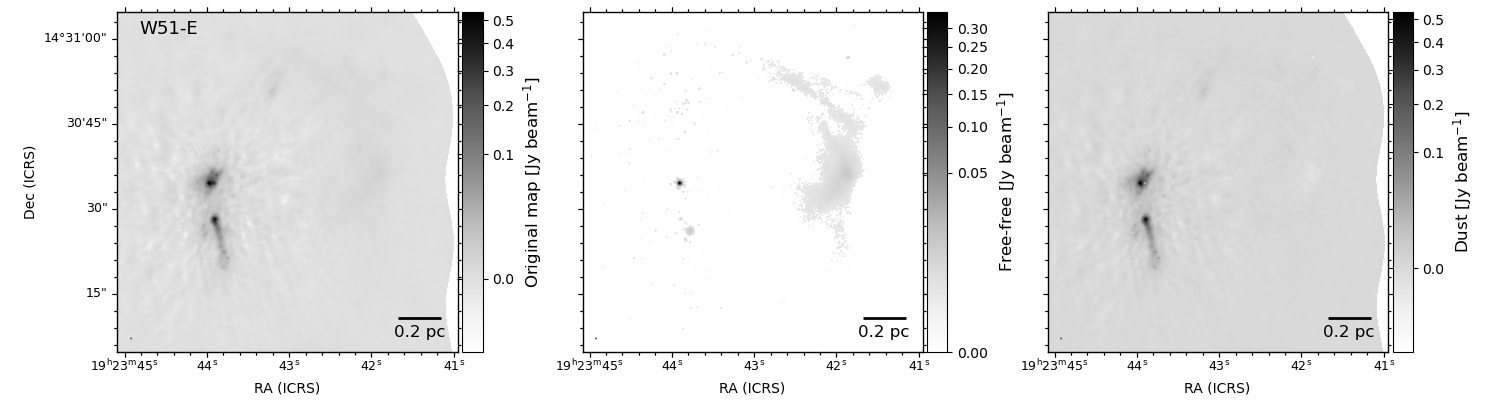}~\\
\includegraphics[width=1.00\linewidth]{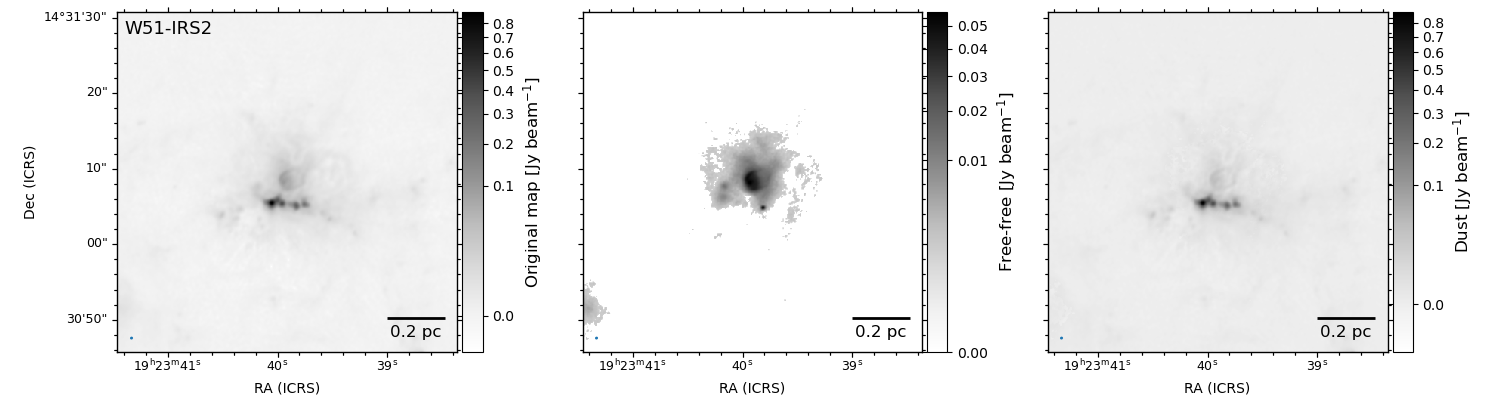}~
\caption{
Continued. 
Comparison of the original 1.3 mm continuum image convolved to the angular resolution of the H$41\alpha$ cube ({\it left} panel), free-free estimation map at 1.3 mm ({\it center}), and the corresponding pure-dust estimation map ({\it left}). The images shown are prior to primary-beam correction. The color normalization is a power law with a 0.33 exponent. The displayed areas are the same as in Fig. \ref{fig:ff-maps}, and correspond to where free-free emission was detected.    
}
\end{figure*}

\section{Filenames and data products} \label{app:products}

We make available the following FITS images in the following url, hosted by the Harvard Dataverse service: \\ \url{https://doi.org/10.7910/DVN/LWILNE}
\begin{itemize}

\item The image cubes containing the hydrogen and helium $41\alpha$ lines, prior to primary-beam correction (see Section \ref{sec:proc}): \texttt{FIELD\_B3\_spw1\_12M\_h41a.JvM.image.contsub.fits}. 
\item 
The regridded free-free estimation $I_{\mathrm{ff},\nu}$ for each protocluster at the standard ALMA-IMF \texttt{cleanest} continuum frequencies of 98.5 and 224.5 GHz. The filenames start with the basename \texttt{FIELD\_B3\_spw1\_12M\_h41a.JvM.image.contsub}, which is standard naming within the ALMA-IMF data reduction pipeline \citep[see][]{Ginsburg22}. The filenames are then followed by \texttt{.m0.ffXXX.XGHz\_Y.Ysigma.regr}, which denotes the fact that they are obtained from a moment 0 map, the frequency of the calculation, the threshold in the moment 0 map above which Eq. \ref{eq:ff_map} is applied, and the applied regridding. Finally, the filenames end with either \texttt{.pbcor.fits} or just with \texttt{.fits}, depending on whether primary-beam correction is applied  or not at the end of the procedure (see Fig. \ref{fig:flowchart}). 
An example filename with PB correction is \texttt{W51-IRS2\_B3\_spw1\_12M\_h41a.JvM.image.contsub.m0.ff224.5GHz\_5.0sigma.regr.pbcor.fits}.  

\item 
The error $\sigma_\mathrm{ff}$ of the free-free estimation. Following the same nomenclature as above but with \texttt{.err} in the filename, e.g., \texttt{W51-IRS2\_B3\_spw1\_12M\_h41a.JvM.image.contsub.m0.ff224.5GHz\_5.0sigma.err.regr.pbcor.fits}.  

\item 
The \texttt{cleanest} continuum image from the release of \citet{Ginsburg22} convolved to the beamsize of the H$41\alpha$ cube. The filename is the same as in the ALMA-IMF data release, but ending with \texttt{.conv.fits}. For example: \\ {\small \texttt{W51-IRS2\_B6\_uid\_\_\_A001\_X1296\_X187\_continuum\_merged\_12M\_robust0\_selfcal9\_finaliter.image.tt0.conv.fits}}

\item 
Estimates of pure dust continuum emission, from the subtraction of the original continuum images minus the free-free estimations, with and without PB correction. The basename of these images is the same as for the original continuum images, but ending with \texttt{.conv.ffsub\_Y.Ysigma.fits}. 
\end{itemize}

\section{Hydrogen and helium $41\alpha$ fitting in the rest of the \textsc{Hii} regions} \label{app:fitting}

Below we show the H$41\alpha$ and He$41\alpha$ fitting in the rest of the \textsc{Hii} regions defined in Section \ref{sec:hiis}, and not shown Fig. \ref{fig:H_He_fitplots}. 

\begin{figure*}
	\includegraphics[width=\textwidth]{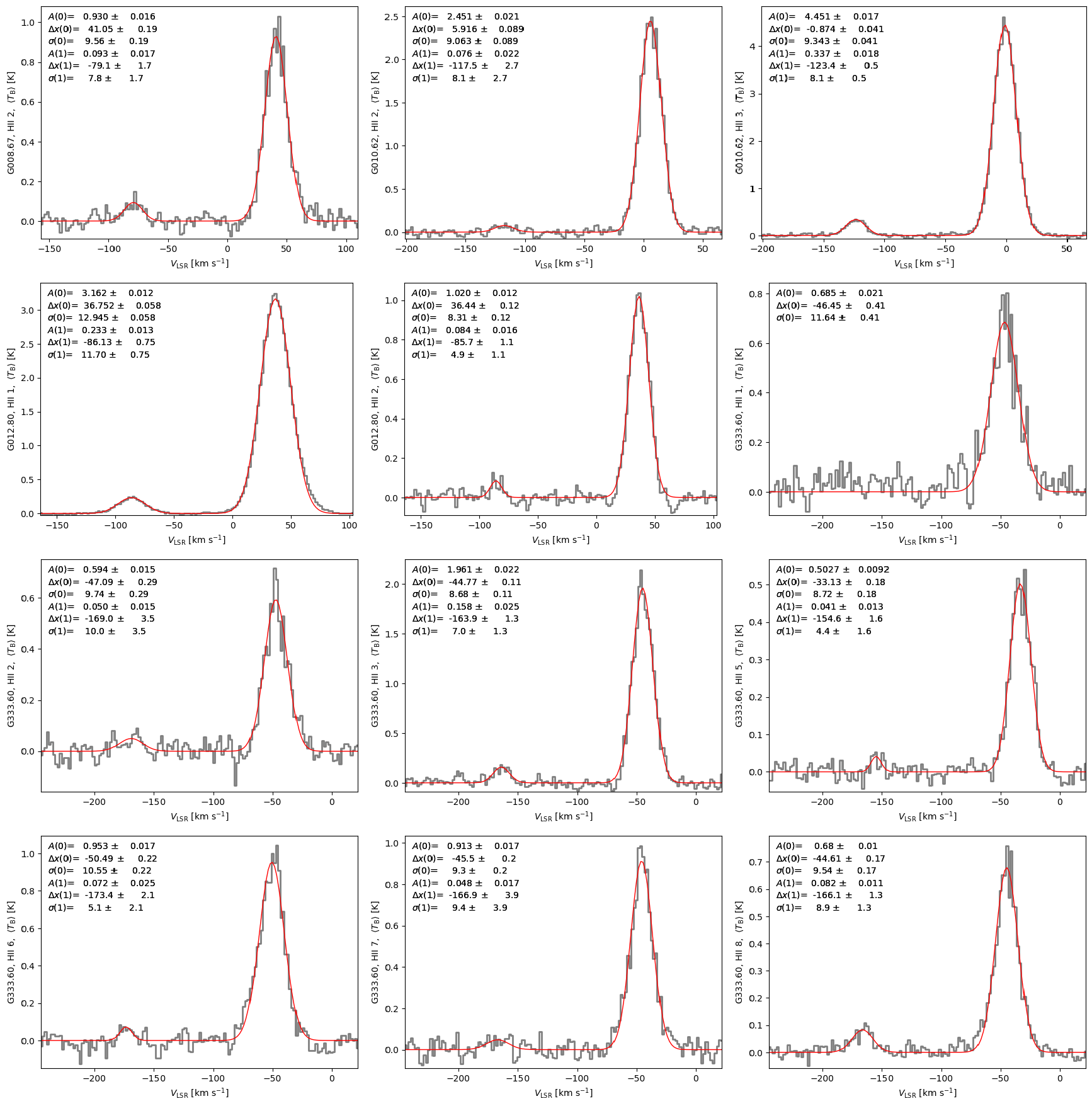}
    \caption{
    Fits to H$41\alpha$ ($\nu_0 = 92.034434$ GHz) and He$41\alpha$ ($\nu_0 = 92.071938$ GHz) recombination line emission in the 12 protoclusters with detection. 
    This figure shows the \textsc{Hii} regions
    not shown in Fig. \ref{fig:H_He_fitplots}, from
    those listed in Table \ref{tab:cont_params}. The fitted parameters and derived abundances are listed in Table \ref{tab:H41_fits}. 
    }
    \label{fig:H_He_fitplots_app1}
\end{figure*}

\addtocounter{figure}{-1}
\begin{figure*}
	\includegraphics[width=\textwidth]{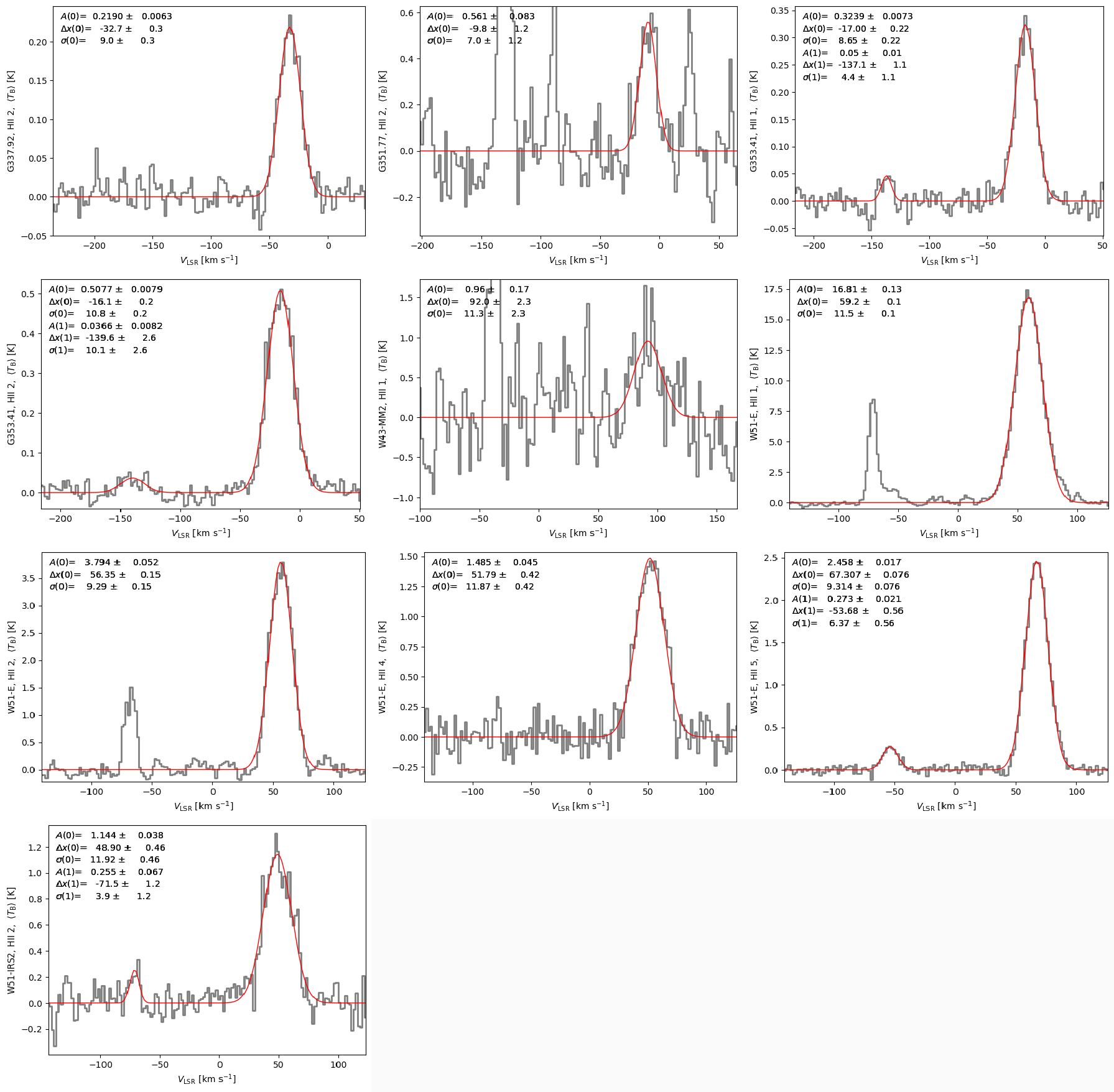}
    \caption{Continued. 
    Fits to H$41\alpha$ ($\nu_0 = 92.034434$ GHz) and He$41\alpha$ ($\nu_0 = 92.071938$ GHz) recombination line emission in the 12 protoclusters with detection. 
    This figure shows the \textsc{Hii} regions
    not shown in Fig. \ref{fig:H_He_fitplots}, from
    those listed in Table \ref{tab:cont_params}. The fitted parameters and derived abundances are listed in Table \ref{tab:H41_fits}. 
    }
\end{figure*}

\section{Abundance and metallicity definitions} \label{app:Y}

In most contexts within astronomy, such as in stellar astrophysics and nucleosynthesis, elemental abundances are given in mass units relative to the total mass $M$ as in: 
\begin{equation}
X + Y + Z = 1, 
\end{equation}

\noindent 
where $X$, $Y$, and $Z$ refer to the hydrogen, helium, and ``metals'' respectively. 

Therefore, the relative mass abundance of helium is given by: 
\begin{equation}
Y = 1 - Z - \frac{m_\mathrm{H} N_\mathrm{H}}{M},     
\end{equation}

\noindent
where $m_\mathrm{H}$ is the mass of atomic hydrogen, $N_\mathrm{H}$ is the total number of hydrogen atoms. 

Rearranging in terms of observables we have: 
\begin{equation}
Y = (1-Z) \biggl( 1 - \frac{1}{1+(N_\mathrm{He} / N_\mathrm{H}) (m_\mathrm{He} / m_\mathrm{H})}   \biggr),     
\end{equation}

\noindent
where the atomic He-to-H mass ratio is $4.002062 / 1.00784 = 3.97147$. 

The solar metallicity is often quoted as $Z_\odot = 0.02$, but recent revisions have taken the value down to $Z_\odot = 0.0153$ \citep{Caffau2011}, a value which is used in modern stellar models \citep[e.g.,][]{Bressan12}. 
Excluding the low- and high-outliers G353.41 and W43-MM2, our typical 
measured values for $N_\mathrm{He} / N_\mathrm{H}$ in the ALMA-IMF \textsc{Hii} regions are in the range from 0.065 to 0.104, with typical errors of $\pm 0.005$. Using $Z_\odot$ from \citet{Caffau2011}, this translates into $Y = 0.202\pm0.012$ to $Y = 0.288\pm0.010$.  
These values are consistent with the inferred value  $Z_\odot/X_\odot = 0.0209$, $Y_\odot = 0.2526$ by \citet{Caffau2011}.
Increasing $Z_\odot$ to 0.02 decreases the inferred values of $Y$ by $\sim 0.001$. 

\bibliography{ALMA-IMF-ff}{}
\bibliographystyle{aasjournal}



\end{document}